\documentclass[fleqn,usenatbib]{mnras}

\usepackage{newtxtext,newtxmath}

\usepackage[T1]{fontenc}

\DeclareRobustCommand{\VAN}[3]{#2}
\let\VANthebibliography\thebibliography
\def\thebibliography{\DeclareRobustCommand{\VAN}[3]{##3}\VANthebibliography}
\usepackage{graphicx}	
\usepackage{amsmath}	
\usepackage{amssymb}	
\usepackage{siunitx}
\usepackage{subcaption}
\usepackage{adjustbox}
\usepackage{tabularx}
\usepackage{multirow, makecell}
\usepackage{booktabs}
\usepackage{hyperref}
\usepackage{enumerate}

\newcommand{\abs}[1]{\left\lvert #1 \right\rvert}
\newcommand{\lya}{Ly$\alpha$}
\newcommand{\pin}{p_{\text{in}}}
\newcommand{\pout}{p_{\text{out}}}
\newcommand{\xlow}{0.304 \pm 0.042}
\newcommand{\xhi}{0.384 \pm 0.133}
\newcommand{\spectre}{{\sc Spectre}}

\title[QSO continua predictions with neural spline flows]{Fully probabilistic quasar continua predictions near Lyman-$\alpha$ with conditional neural spline flows}

\author[D. M. Reiman et al.]{
David M. Reiman$^{1}$\thanks{E-mail: dreiman@ucsc.edu},
John Tamanas$^{1}$,
J. Xavier Prochaska$^{2,3}$,
and Dominika \v{D}urov\v{c}\'{i}kov\'{a}$^{4}$
\\
$^{1}$Department of Physics, University of California Santa Cruz, 1156 High Street, Santa Cruz, California 95064, USA \\
$^{2}$Department of Astronomy \& Astrophysics, University of California Santa Cruz, 1156 High Street, Santa Cruz, California 95064, USA \\
$^{3}$Kavli IPMU, the University of Tokyo (WPI), Kashiwa 277-8583, Japan \\
$^{4}$New College, University of Oxford, Holywell Street, Oxford OX1 3BN, UK
}

\date{Accepted XXX. Received YYY; in original form ZZZ}

\pubyear{2020}

\begin{document}
\label{firstpage}
\pagerange{\pageref{firstpage}--\pageref{lastpage}}
\maketitle

\begin{abstract}
Measurement of the red damping wing of neutral hydrogen in quasar spectra provides a probe of the epoch of reionization in the early Universe. Such quantification requires precise and unbiased estimates of the intrinsic continua near Lyman-$\alpha$ (Ly$\alpha$), a challenging task given the highly variable Ly$\alpha$ emission profiles of quasars. Here, we introduce a fully probabilistic approach to intrinsic continua prediction. We frame the problem as a conditional density estimation task and explicitly model the distribution over plausible blue-side continua ($\SI{1190}{\angstrom} \leq \lambda_{\text{rest}} < \SI{1290}{\angstrom}$) conditional on the red-side spectrum ($\SI{1290}{\angstrom} \leq \lambda_{\text{rest}} < \SI{2900}{\angstrom}$) using \textit{normalizing flows}. Our approach achieves state-of-the-art precision and accuracy, allows for sampling one thousand plausible continua in less than a tenth of a second, and can natively provide confidence intervals on the blue-side continua via Monte Carlo sampling. We measure the damping wing effect in two $z>7$ quasars and estimate the volume-averaged neutral fraction of hydrogen from each, finding  $\bar{x}_\text{HI}=\xlow$ for ULAS J1120+0641 ($z=7.09$) and $\bar{x}_\text{HI}=\xhi$ for ULAS J1342+0928 ($z=7.54$).
\end{abstract}

\begin{keywords}
 intergalactic medium -- quasars: general -- quasars: emission lines -- dark ages, reionization, first stars
\end{keywords}


\begin{figure*}
    \centering
	\includegraphics[width=\textwidth]{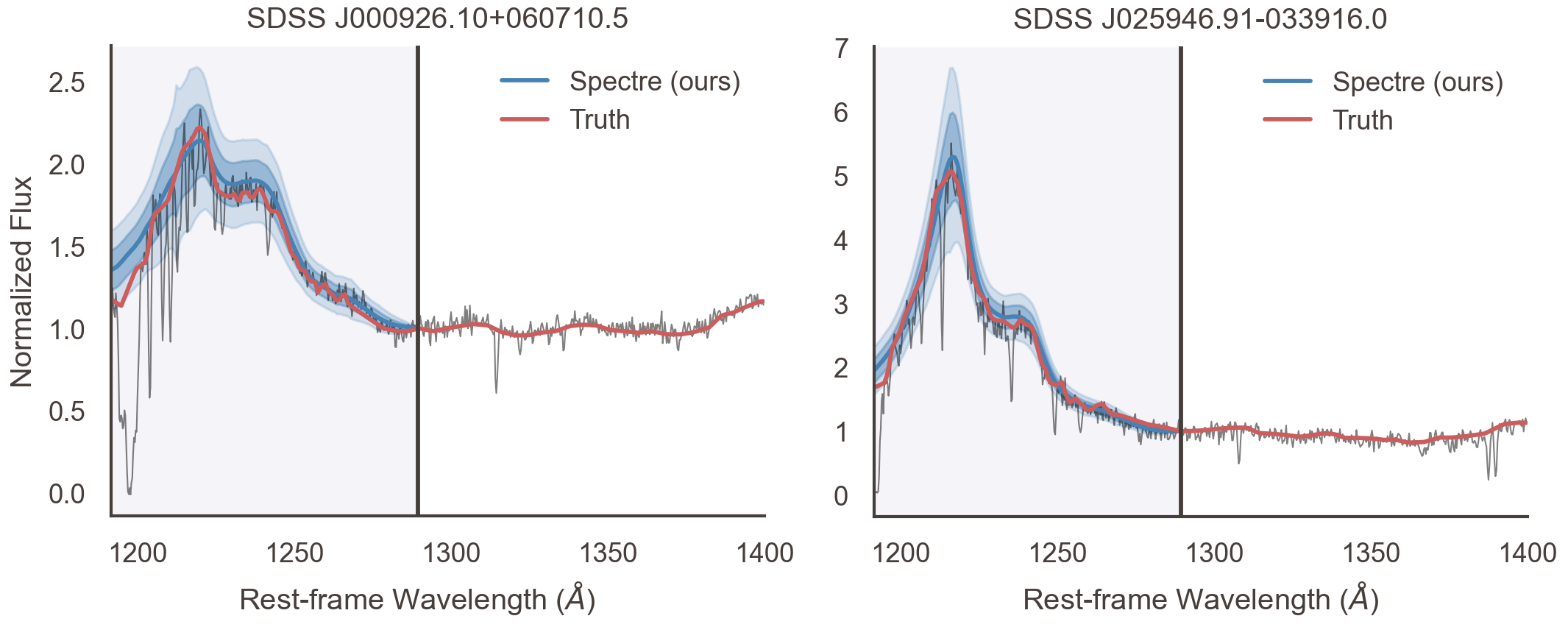}
    \caption{To infer the blue-side continuum, \spectre\ samples one thousand predictions conditional on the red-side. The assumed truth, shown in red, is the smoothed continua estimation from our preprocessing scheme, and the raw flux is shown in grey. }
\end{figure*}

\section{Introduction}

Prior to the emergence of the first luminous sources, the intergalactic medium (IGM) was filled with a dense gas of neutral hydrogen. During the epoch of reionization, nascent stars, galaxies and quasars ionized the surrounding IGM with ultraviolet (UV) radiation. Reionization induced a patchy topology upon the universe wherein ionization bubbles grew about each source until the ionization regions merged and percolated, and residual neutral hydrogen was left primarily in the deep potential wells of dark matter halos. Recent measurements of the cosmic microwave background suggest that reionization of the IGM 
occurred at $z \sim 8$ \citep{planck2018}, 
but there remains uncertainty about its precise timing and nature---constraining the history of reionization is a goal of modern cosmology.

A largely neutral IGM leaves marked signatures in source spectra. At redshifts beyond $z \approx 6$, the spectra of sources sufficiently luminous to be observed with modern telescopes exhibit a near-complete suppression of flux blueward of Ly$\alpha$---an effect known as the Gunn-Peterson trough \citep{gunnPeterson}. In comparison, at lower redshifts where the residual neutral hydrogen has gathered in the potential wells of dark matter halos and other large-scale structures, we observe discrete absorption features (the Ly$\alpha$ forest) in source spectra that trace out the matter field of the universe \citep[e.g.][]{lymanAlphaForest1, lymanAlphaForest2}. These latter observations confirm the presence of a highly ionized IGM.

These considerations suggest that one could probe the epoch of reionization by searching for the presence of the Gunn-Peterson trough in a sequence of luminous sources of increasing redshift. However, the presence of the Gunn-Peterson trough alone is insufficient to prove that the emitting source is surrounded by a neutral IGM \citep{dampingWing}---due to the considerable optical depth of the neutral IGM, even a modest residual neutral fraction of hydrogen in a largely reionized IGM will suppress the transmittance of flux blueward of Ly$\alpha$ to near-zero. 

Instead, unambiguous evidence of a neutral IGM may be provided by measurement of the \textit{red damping wing} of the Gunn-Peterson trough \citep{dampingWing}, which is only identifiable for very large column densities of neutral hydrogen such as those at or prior to reionization. This broad absorption feature extends redward of Ly$\alpha$ out to a rest-frame wavelength of $\lambda_{\text{rest}} \approx \SI{1260}{\angstrom}$ in the quasar's rest-frame.  By measuring the damping wing in a series of high-redshift sources, one can place constraints on the timeline of the early universe IGM phase transition from neutral to reionized. Thus far, the only sources luminous enough to permit measurement of the damping wing are quasars, although there remains optimism that gamma-ray burst afterglows may offer an additional avenue \citep{grb}.

Measurement of the damping wing is complicated by a variety of factors, notably: (i) the possibility of a potent absorption system along the quasar line-of-sight whose proximity alters the profile of the wing, and (ii) the uncertainty in our estimation of the intrinsic quasar flux (termed the ``continuum'') whose profile allows us to measure the damping wing width. As noted by \citet{dampingWing}, continua prediction is especially difficult when the emitting source is a quasar since ``quasars have strong, broad Ly$\alpha$ emission lines with profiles that are highly variable.'' Thus, a model which predicts intrinsic quasar continua should be expressive and ideally provide calibrated uncertainties with its predictions.

Another complicating factor is estimating the extent of the quasar near-zone. In the proximity of powerful UV sources such as quasars, the increased photo-ionization rate of neutral hydrogen leads to a diminishing number of Ly$\alpha$ absorbers near the redshift of the source. This consequence is known as the line-of-sight proximity effect \citep{proximityEffect}, and its associated spectral feature is denoted the \textit{ionization near-zone}, or simply the proximity zone. Our model of the damping wing relies on our knowledge of the blueward edge of the quasar proximity zone, and though we have heuristics to locate the edge of the near-zone, our uncertainty in the true location affects our estimates of the damping wing strength.

Intrinsic continua prediction has been studied widely and approached in a variety of manners. Previous approaches generally predict the blue-side continua ($\SI{1190}{\angstrom} \leq \lambda_{\text{rest}} < \SI{1290}{\angstrom}$) using information encoded in the red-side spectrum ($\lambda_{\text{rest}} \geq \SI{1290}{\angstrom}$). This information arises from correlations in the emission features of the blue and red-side spectra \citep{corr1}. Such models are typically trained on the spectra of quasars at moderate redshift---a typical redshift range is $z \in [2.1,\  2.5]$ \citep{greigLya}---which cover the Ly$\alpha$ and Mg \textsc{ii} broad emission lines. These lines strongly constrain the standard pipeline estimates for quasar redshifts \citep{dr14}.
Since such redshift estimates are a major source of bias in any model which aims to impute spectra, selecting a redshift range which includes such emission features limits our exposure to strong systematics. Approaches such as these often involve a pair of models: the \textit{primary model} which infers the blue-side continua given the red-side spectrum, and the \textit{secondary model} which probes the similarity of high-redshift quasars (our targets for inference) to our moderate-redshift training set.

The primary model predicts the intrinsic (i.e. unabsorbed) continua near Ly$\alpha$ given the redward spectrum. Primary models are often fundamentally non-probabilistic (with a notable exception being the fully Bayesian framework of \citealt{greigLya}) though many employ methods such as ensembling or prediction on nearest neighbors to approximate confidence intervals during inference. For example, previous approaches have employed principal component analysis (PCA) to the blue and red sides of the spectrum and subsequently related the coefficients using multiple linear regression \citep{suzukiPCA,daviesPCA} or an ensemble of neural networks \citep{qsanndra}.

The secondary model probes the similarity in spectral characteristics between the moderate-$z$ training set and the high-z quasars which are targets for constraining the epoch of reionization. High similarity assures us that the high-z targets are \textit{in-distribution} and therefore valid inputs to our primary model. Here, similarity is tantamount to likelihood though typically simple models or proxies are used. A number of approaches have been chosen in related studies, for example the reconstruction error of an autoencoder \citep{qsanndra} or the likelihood of PCA coefficients under a Gaussian mixture model \citep{daviesPCA}.

In this article we introduce \spectre, a fully probabilistic approach to intrinsic continua prediction which utilizes normalizing flows as both the primary and secondary models. We frame the problem as one of conditional density estimation: what is the probability distribution over blue-side continua given the redward spectrum? In doing so, \spectre\ achieves state-of-the-art precision, allows for sampling one thousand plausible continua in less than a tenth of a second on modern GPUs, and can natively provide confidence intervals on the blue-side continua via Monte Carlo sampling. In addition, our secondary model provides the likelihood ratio as a background contrastive score used to measure how in-distribution a given quasar continuum lies, offering a new perspective on the probability of high-redshift quasar spectra under a generative model trained on moderate-redshift quasars. Both primary and secondary models are applied to continua from high redshift ($z>7$) quasars, ULAS J1120+0641 ($z=7.09$) \citep{J1120} and ULAS J1342+0928 ($z=7.54$) \citep{J1342}, in order to infer the neutral hydrogen fraction of the universe during the epoch of reionization.

This manuscript is organized as follows: \S\ref{sec:related_work} provides a brief overview of previous approaches to intrinsic continua prediction. In \S\ref{sec:background} we introduce normalizing flows, describe their usefulness in scientific applications, and provide a brief introduction to the particular variety we employ in this work: neural spline flows. Then, in \S\ref{sec:methods} we detail our approach by first outlining our data preprocessing scheme, describing our flow-based model and explaining our measurement of the damping wing in quasar spectra. Results are presented in \S\ref{sec:results} for our constraints on the reionization history of the early universe from measuring the damping wing in two $z > 7$ quasars. In \S\ref{sec:conclusion}, we conclude with a summary of our work and further discuss the use of flows for probabilistic modeling in the sciences.


\section{Related Work}
\label{sec:related_work}

Previous approaches to intrinsic quasar continua prediction have ranged from fully Bayesian models operating purely upon emission features \citep{greigLya} to full-spectrum principal component analyses on blue/red-side continua to learn correlations between the dominant modes of variation in each, as in \citet{daviesPCA}; \citet{qsanndra}. In these approaches, the authors fit models on moderate redshift quasars (typically $z \approx 2$) and apply them to quasars near or at reionization ($z \approx 7$).

In the Bayesian approach of \citet{greigLya}, the authors construct a covariance matrix describing the correlations of high-ionization emission line features in moderate redshift quasar spectra. The emission line profiles are compressed into three features: line width, peak height and velocity offset. Each of their chosen emission features (\lya, Si \textsc{iv}/O \textsc{iv}, C \textsc{iv}, and C \textsc{iii}) are modeled with either a single or double component Gaussian profile. The Gaussian components are fit to each spectrum in their training set using a Markov Chain Monte Carlo approach with a $\chi^2$ likelihood function. After fitting each element of their training set, the authors compute the correlation matrix between the emission line features for all included lines. For reconstructing the \lya\ emission of high-redshift quasars, the red-side emission features are fit in a manner identical to the training set. The likelihood of its parameter vector is modeled via a high-dimensional Gaussian with mean equal to the mean parameter vector of the training set and covariance equal to the covariance matrix computed from the training set. By doing so, the authors assume that the marginal distributions of each parameter can be described by a Gaussian. Reconstruction of the blue-side continua then proceeds by collapsing the likelihood function along all dimensions corresponding to the parameters of the red-side emission features, leaving only the 6-dimensional conditional likelihood of the \lya\ emission: three parameters for each of its two Gaussian components. Here, a prior on the blue-side emission features is introduced by fitting on unabsorbed quasar spectra at $z \approx 6$. The resulting posterior is a joint distribution over these six parameters which provides a probabilistic model for the blue-side intrinsic continua conditioned on the red-side emission features. 

Subsequently, the authors published analyses on the damping wing of hydrogen in two available $z > 7$ QSOs \citep{greigConstraints1, greigConstraints2} using large-scale epoch of reionization simulations. To do this, they sampled ${\sim}10^5$ plausible blue-side continua from their model and multiplied each by ${\sim}10^5$ synthetic damping wing opacities from their simulations of the epoch of reionization. These ${\sim}10^{10}$ mock spectra were compared to the observed spectrum of the high-redshift QSOs using a $\chi^2$ likelihood function. The final estimate of the neutral fraction was then calculated by weighting the neutral fraction of each synthetic damping wing by its marginal likelihood (over all mock spectra) with reference to the observed spectrum and computing the weighted average.

Other work makes use of principal component analysis (PCA) to reduce the dimensionality of the problem. PCA produces a set of linearly uncorrelated PCA eigenvectors. A spectrum can then be reconstructed with a sum of PCA eigenvectors multiplied by the appropriate coefficients. Among techniques which employ PCA, the number of coefficients may be chosen by hand or to satisfy some criterion on the explained variance (e.g. 99\%). Correlations between blue and red-side coefficients are then modeled with either a linear model  \citep[e.g.][]{suzukiPCA, ParisPCA, EilersPCA, EilersPCA2, daviesPCA} or an ensemble of neural networks \citep{qsanndra}. Inference on high-redshift quasars is then performed by encoding the the red-side spectrum into its PCA coefficients and using the trained model to predict the blue-side coefficients. Using the blue-side coefficients and PCA eigenvectors, the blue-side continua can be reconstructed and used as a prediction of the intrinsic continua. The neutral fraction of hydrogen is then estimated using either a simplified model of the red damping wing (as in \citealt{qsanndra}) or via full hydrodynamical modeling of the neutral IGM (as in \citealt{daviesX}).


\section{Background}
\label{sec:background}

This section describes normalizing flows in moderate detail. For an exhaustive introduction and tutorial refer to \citet{normFlowsReview}---we will adopt a similar notation from here onwards: $\mathbf{x}$ is a D-dimensional\footnote{The dimensionality D is defined by the data---for spectroscopy, D is determined by the resolution and wavelength range of the spectroscope, although D may change due to subsequent preprocessing.} random vector (e.g. the measured spectrum of a quasar) generated from an underlying distribution $p_x^*(\mathbf{x})$, and $\mathbf{u} \sim p_u(\mathbf{u})$ is the latent (or hidden) representation of $\mathbf{x}$ in a D-dimensional isotropic Gaussian space. The representations are related via a transformation $T \colon \mathbb{R}^D \to \mathbb{R}^D$ such that $\mathbf{x} = T(\mathbf{u})$. We model the true distribution over $\mathbf{x}$ via a distribution $p_x(\mathbf{x};\ \boldsymbol{\theta})$ produced by a neural network with parameters $\boldsymbol{\theta}$.

\subsection{Normalizing Flows}
\label{sec:flows}

Normalizing flows model complex probability densities by mapping samples between a base density and the distribution of interest, i.e. the data distribution. The transformation, $T$, is composed of a series of invertible mappings $T_i$, or \textit{bijections}, each of which must be differentiable. The base density is commonly chosen to be Gaussian, with the requirement that its dimensionality be equal to the dimensionality of the data to maintain invertibility. Since the composition of differentiable, invertible mappings $T = T_0 \circ T_1 \circ ... \circ T_N$ is itself differentiable and invertible, the normalizing flow acts as a diffeomorphism between the data space and the Gaussian latent space. In our application, the flow then learns a bijective mapping between Gaussian-distributed latent samples and blue-side quasar continua (conditional on the red-side spectrum).

The density of a random vector in the data space is then well-defined and easily computed via a change of variables---we simply cast our data into the Gaussian latent space where density evaluation is trivial. For a datum $\mathbf{x}$ drawn from a \textit{D}-dimensional target distribution $p^*_x$, we can model its density in the following manner:

\begin{equation}
    p_x(\mathbf{x}) = p_u(\mathbf{u})\abs{\det{J_T(\mathbf{u})}} ^{-1}
	\label{eq:change-of-variables}
\end{equation}
 where $\mathbf{x} = T(\mathbf{u})$, $p_u$ is the base density and $J_T$ is the Jacobian of the transformation $T$ which maps samples from the Gaussian space to the data space.
 
Sampling is similarly uncomplicated: latent vectors are sampled in the Gaussian space and transformed into data space samples via $T$. Quantities such as confidence intervals can then easily be computed via Monte Carlo sampling and are generally calculable from a single (large batch) forward pass provided the data dimensionality and model size are modest.

In practice, we employ neural networks to parameterize our invertible transformations $T_i$ (which together compose $T$) and cleverly choose the form of the transformations such that the Jacobian is lower triangular. The neural networks themselves are parameterized by a parameter vector $\boldsymbol{\theta}$. Since the determinant of a lower triangular matrix is simply the product of its diagonal elements, this reduces the complexity of the determinant calculation from $\mathcal{O}(D^3)$ to $\mathcal{O}(D)$.

To train such a model, we minimize the divergence between the target distribution $p^*_x(\mathbf{x})$ and the distribution parameterized by the flow, $p_x(\mathbf{x}; \boldsymbol{\theta})$. A common choice of divergence is the Kullback-Leibler (KL) divergence which measures the loss of information when using the model distribution to estimate the target distribution.

\begin{align}
    \mathcal{D}_{KL}\ [p^*_x(\mathbf{x})\ \Vert\ p_x(\mathbf{x}; \boldsymbol{\theta})] &= \mathbb{E}_{p^*_x(\mathbf{x})}\Big[\log{\frac{p^*_x(\mathbf{x})}{p_x(\mathbf{x}; \boldsymbol{\theta})}}\Big] \\
    &= \mathbb{E}_{p^*_x(\mathbf{x})}\Big[\log{p^*_x(\mathbf{x}) - \log{p_x(\mathbf{x}; \boldsymbol{\theta})}}\Big] \\
    &= - \mathbb{E}_{p^*_x(\mathbf{x})}\Big[\log{p_x(\mathbf{x}; \boldsymbol{\theta})}\Big] + \text{const.} \\
	\label{eq:kl-divergence}
\end{align}
where $\mathcal{D}_{KL}$ is the KL-divergence and $\mathbb{E}_{p^*_x(\mathbf{x})}$ denotes the expectation with respect to the target distribution $p^*_x(\mathbf{x})$.

We can then identify an appropriate loss function for training by writing the model density in terms of the flow transformation and its Jacobian, using the fact that the determinant of the inverse of an invertible transformation is the inverse of the determinant of the transformation and recalling that $\mathbf{u} = T^{-1}(\mathbf{x})$. 

\begin{align}
    \mathcal{L}(\boldsymbol{\theta}) = - \mathbb{E}_{p^*_x(\mathbf{x})}\Big[\log{p_u(T^{-1}(\mathbf{x}; \boldsymbol{\theta}))} + \log{\abs{\det{J_{T^{-1}}(\mathbf{x}; \boldsymbol{\theta})}}}\Big] \\
	\label{eq:loss}
\end{align}

Given a dataset of samples from the target distribution $\{\mathbf{x}_n\}_{n=1}^N$ (e.g. a collection of quasar spectra) we can approximate the expectation over $p^*_x(\mathbf{x})$ via Monte Carlo.

\begin{align}
    \mathcal{L}(\boldsymbol{\theta}) \approx - \frac{1}{N}\sum_{n=1}^N\Big[\log{p_u(T^{-1}(\mathbf{x}_n; \boldsymbol{\theta}))} + \log{\abs{\det{J_{T^{-1}}(\mathbf{x}_n; \boldsymbol{\theta})}}}\Big] \\
	\label{eq:loss-monte-carlo}
\end{align}

Thus, training a normalizing flow amounts to explicitly maximizing the likelihood of our dataset where the likelihood of each datum is exactly calculable by casting it into a Gaussian latent space where density calculations are trivial. The only caveat is that we must compute the determinant of the Jacobian of the transformation relating the data space to the Gaussian latent space. 

The ability to do exact density evaluation make normalizing flows an attractive model for probabilistic modeling. Indeed, deep generative models which admit exact likelihoods are rare: variational autoencoders (VAEs, see \citealt{vae}) admit only approximate likelihoods and generative adversarial networks (GANs, see \citealt{gan}) admit no likelihoods at all. Autoregressive generative models \citep{pixelrnn, pixelcnn, transformer} offer exact density evaluation but generate samples via ancestral sampling which requires repeated forward passes through the network. In contrast, normalizing flows can be designed to offer both density estimation and sampling in a single forward pass.

\subsection{Transforms}

\begin{figure*}
    \centering
    \begin{subfigure}{\columnwidth}
    	\includegraphics[height=5.2cm]{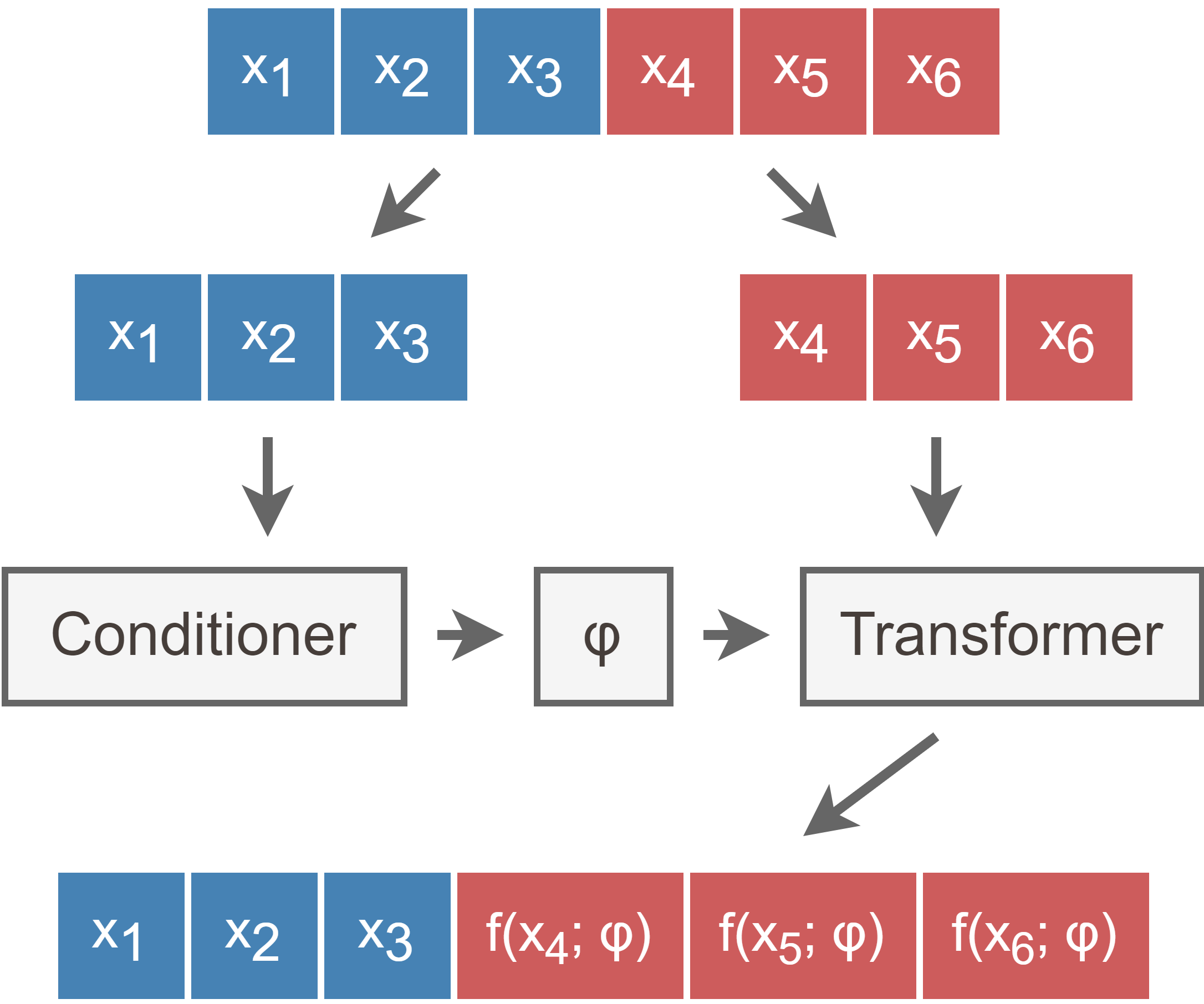}
    \end{subfigure}
    \begin{subfigure}{\columnwidth}
    	\includegraphics[height=5.2cm]{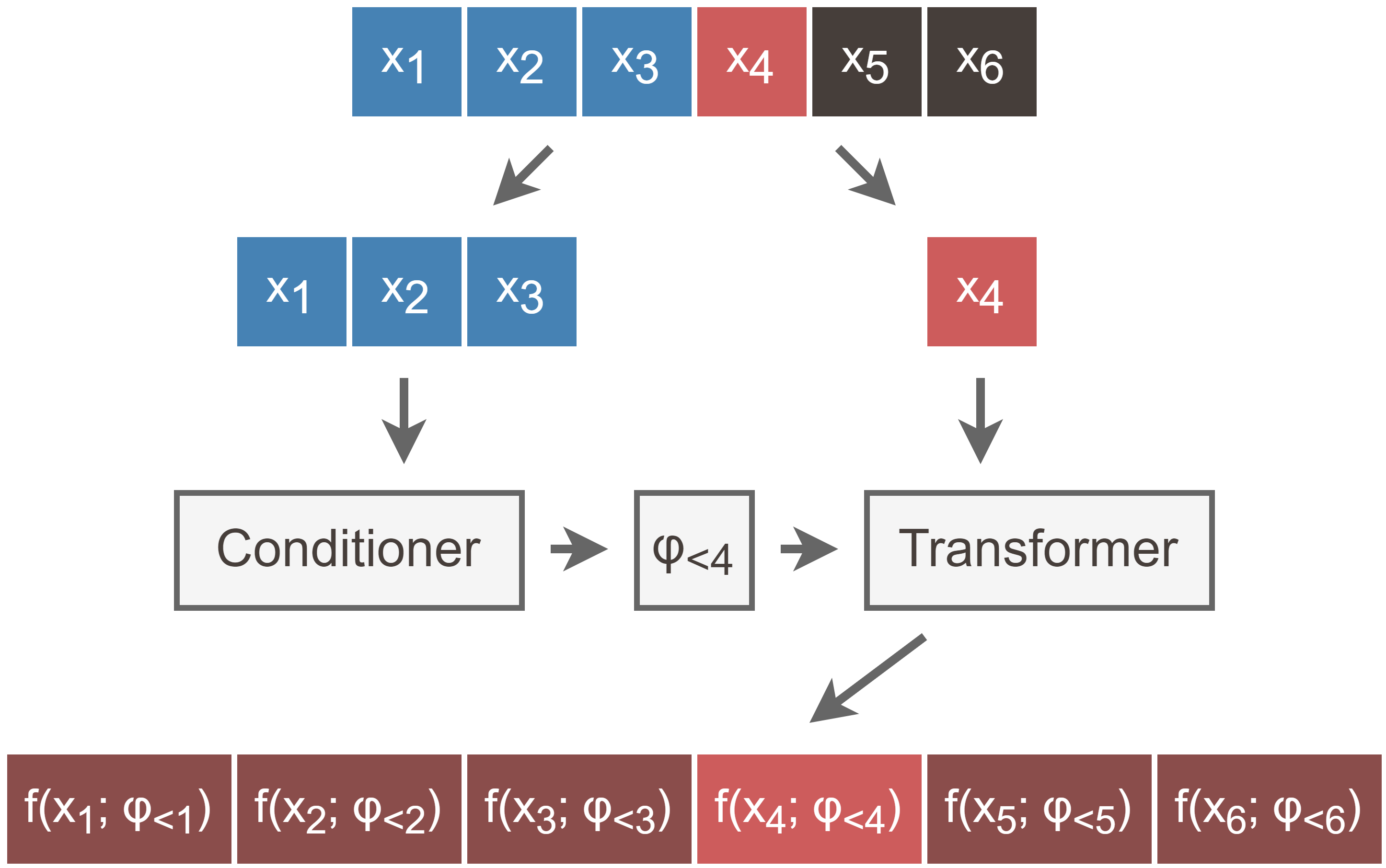}
    \end{subfigure}
    \caption{\textbf{Left:} Example coupling layer transform for a $D=6$ dimensional input vector, $\mathbf{x}_{1:6}$. Upon splitting the input features into halves, the first half is consumed by the conditioner which outputs a parameter vector, $\boldsymbol{\phi}(\mathbf{x}_{1:3})$. This vector parameterizes a monotonic transformation of the second half features, $\mathbf{x}_{4:6}$. The coupling layer output is then given by the concatenation of the identity features with the transmuted features. \textbf{Right:} Example autoregressive layer transform for the same input vector. The input is split into $D=6$ pieces. Features at each dimension index $i$ undergo a mapping parameterized by all features with dimension index lower than $i$. Here, we display the process only for element $x_4$ and use shorthand notation such that $\boldsymbol{\phi}_{<i} = \boldsymbol{\phi}(\mathbf{x}_{<i})$.}
    \label{fig:flow_transforms}
\end{figure*}

Though the mathematical underpinning of normalizing flows is elegant, their practical application is limited by the calculation of a determinant for each bijection $T_i$ during each forward pass. To circumvent this, most flow models employ transformations designed to yield lower triangular Jacobians.  Since the determinant of a lower triangular matrix is easily computed by multiplying its diagonal elements, the complexity of the determinant computation then scales linearly in the data dimensionality, $D$. Two such choices of transformation are the \textit{coupling transform} of \citet{nice}; \citet{realNVP} and the \textit{autoregressive transforms} of \citet{iaf}; \citet{maf}.
We discuss the merits of each in turn.

\subsubsection{Coupling Transforms}

Coupling transforms operate by dividing an input datum into halves, then using the former half (hereafter the \textit{identity features}) to predict the parameters of an invertible transformation on the latter half (hereafter the \textit{transmuted features}). Invertibility is enforced by restricting our transformations to be strictly monotonic. The identity features remain untransformed as indicated by their name. After each coupling layer, the dimensions of the data are randomly permuted (imposing an arbitrary ordering at the next layer) to allow features of each data dimension an opportunity to be transformed at some layer of the flow. Note that permutations themselves are invertible transformations with a determinant of $1$ or $-1$. The general form of a coupling transform $T$ is shown below (and a diagram is provided in Fig.~\ref{fig:flow_transforms}, left).

\begin{align}
    \mathbf{y}_{1:d} &= \mathbf{x}_{1:d} \\
    \mathbf{y}_{d+1:D} &= f(\mathbf{x}_{d+1:D};\ \boldsymbol{\phi}(\mathbf{x}_{1:d}))
\end{align}
In the normalizing flow literature, $\boldsymbol{\phi}$ is referred to as the \textit{conditioner} and $f$ as the \textit{transformer}. The conditioner is typically an arbitrary neural network and the transformer is any strictly monotonic function.

Commonly, neural networks in a coupling layer use the identity features to parameterize an affine transformation on the transmuted features. In such cases, the transformer $\boldsymbol{\phi}$ outputs a set of scale and bias parameters which then act on the latter half of the input.

\begin{align}
    \boldsymbol{\phi}(\mathbf{x}_{1:d}) = \{\boldsymbol{\alpha}(\mathbf{x}_{1:d}),\  \boldsymbol{\beta}(\mathbf{x}_{1:d})\} \\
    f(\mathbf{x}_{d+1:D};\ \boldsymbol{\phi}(\mathbf{x}_{1:d})) = \boldsymbol{\alpha}(\mathbf{x}_{1:d}) \cdot \mathbf{x}_{d+1:D} + \boldsymbol{\beta}(\mathbf{x}_{1:d})
\end{align}

Flows employing such transformations have produced promising results in practice but often require an immense number of coupling layers (often hundreds) to model complicated and high-dimensional probability distributions such as those over natural images \citep{glow}.

Promising recent approaches \citep{polynomialFlows, neuralSplines} in which the identity features are used to predict the parameters of a monotonically increasing piecewise spline have been shown capable of modeling highly multimodal distributions with state-of-the-art results (for flows) in log-likelihood scores. We will make use of such a flow in this work.

Coupling layers can also be made conditional in many ways. Since the conditioner is typically an arbitrary neural network, the output of this network can be conditioned on any additional information by, for example, concatenating the conditioning information onto the identity features before predicting the transformation parameters. Given $F$-dimensional conditioning information $c_{1:F}$, the transformation parameters are then computed as $\boldsymbol{\phi}(\mathbf{x}_{1:d}\cdot\mathbf{c}_{1:F})$ where $\cdot$ is the concatenation operator. Generally, each layer of the flow would be conditioned in this manner.

\subsubsection{Autoregressive Transforms}

Autoregressive transforms (see Fig.~\ref{fig:flow_transforms}, right) enforce a lower triangular Jacobian by specifying the following form for their transforms:

\begin{equation}
    \mathbf{y}_{i} = f(\mathbf{x}_{i};\ \boldsymbol{\phi}(\mathbf{x}_{<i}))
\end{equation}

With $\boldsymbol{\phi}$ as the \textit{conditioner} and $f$ as the \textit{transformer}. To make this an invertible transformation, the transformer is again chosen to be a monotonic function of $\mathbf{x}_i$. If the transformer and conditioner are flexible enough to represent any function arbitrarily well (as neural networks are), then autoregressive flows are able to approximate any distribution arbitrarily well (see \citealt{normFlowsReview}).

Autoregressive flows can have either one-pass sampling and $D$-pass density estimation or $D$-pass sampling and one-pass density estimation \citep{maf, iaf}. Because flows are usually trained by maximizing the likelihood of the data with respect to model parameters, autoregressive flows are commonly chosen for one-pass density estimation. Autoregressive transforms can also be made conditional by manner similar to coupling transforms: conditional features, $c_{1:F}$, are concatenated onto $\mathbf{x}_{<i}$ before being inputted to the transformer.

\subsubsection{Choosing a Transform}

The choice of transform depends on the task at hand. If one is only interested in density estimation, autoregressive flows are typically chosen because of their capacity to approximate any distribution arbitrarily well with fewer layers than their coupling transform counterparts. If one would like to sample from the model, however, it is often preferable to choose a coupling transform because it offers one-pass density estimation for maximum likelihood training and one-pass data generation. Though, if efficient sampling is the only criterion, inverse autoregressive flows \citep{iaf} are also an option. Recent work has demonstrated how to model distributions of data which are invariant to transformations of a given symmetry group using equivariant coupling layers \citet{qcdFlows}. In such a case, using coupling layers for density estimation may provide a better inductive bias since the coupling layer can encode the symmetry.

In this paper, we make use of both kinds of transforms: coupling in the primary model for efficient sampling of blue-side continua, and autoregressive in the secondary model for density estimation.

\subsection{Neural Spline Flows}

In contrast to affine flows where the conditioner produces the parameters of a strictly linear (and thereby inflexible) mapping, neural spline flow conditioners parameterize a piecewise spline which can approximate any differentiable monotonic function in the spline region. The added expressivity of spline layers allow neural spline flows to model complex, multi-modal probability densities with significantly  fewer neural network parameters than their affine equivalent.

Neural spline flows make use of the identity features to predict the parameters of a piecewise spline in a region $x,\ y\in[-B,\ B]$ (hereafter the \textit{spline region}) where $B$ is a hyperparameter. The spline is required to be strictly monotonic such that the mapping is one-to-one and thereby invertible. The transformation is piecewise-defined in $K$ different bins spanning the spline region and is linear ($y=x$) beyond. The $K+1$ bin edges are referred to as \textit{knots}. 

Polynomial families of functions are often chosen for the spline---originally up to and including degree two polynomials \citep{polynomialFlows} and subsequently up to and including degree three \citep{cubicFlows}. 
Recently, \citet{neuralSplines} introduced flows which employ rational quadratic splines: a family of functions defined by the division of two quadratic functions. These functions are highly expressive and yet simple to invert. Flows which make use of such transforms are referred to by \citet{neuralSplines} as \textit{rational quadratic neural spline flows} (RQ-NSF).

In RQ-NSFs, each of the $K$ bins are assigned monotonic rational quadratic functions parameterized by a neural network. In total, $3K-1$ parameters define a piecewise rational quadratic spline with $K$ bins for a single data dimension: $K$ bin widths (size in x), $K$ bin heights (size in y) and $K-1$ derivatives at the $K-1$ internal knots (since the derivative at the two outer knots must be unity).

For bin $k$, defining the bin width as $\Delta x_k = x_{k+1} - x_{k}$, the bin height as $\Delta y_k = y_{k+1} - y_{k}$, the derivative at knot $k$ as $\delta_k$, the constant $s_k = \Delta y_k / \Delta x_k$ and function $\xi(x) = (x - x_k) / \Delta x_k$, the rational quadratic spline is then defined as follows.

\begin{align}
    r_k(\xi) = \frac{\alpha_k(\xi)}{\beta_k(\xi)} = y_k + \frac{\Delta y_k\big[s_k\xi^2 + \delta_k\xi(1-\xi)\big]}{s_k + \big[\delta_{k+1}+\delta_{k}-2s_k\big]\xi(1-\xi)}
    \label{eq:rational-quadratic-spline}
\end{align}

It should be noted that the transformation acts elementwise---a unique spline is parameterized for each dimension of the transmuted features. Thus, for $i = d+1,\ d+2,\ \ldots,\ D$ (indexing the transmuted features) we parameterize a spline $r_k^i$ such that $y_i = r_k^i(x_i)$. Then, the determinant of the Jacobian of the coupling transformation can be written as follows.

\begin{align}
    \det{J_{T}} = \det{\frac{\partial T_i}{\partial x_j}} = \prod_{i=1}^d 1 \times \prod_{i=d+1}^D \frac{\partial f_i}{\partial x_i} = \prod_{i=d+1}^D \frac{\partial r_k^i}{\partial x_i}
\end{align}

Where the derivative of the spline is shown below.

\begin{align}
    \frac{dr_k}{dx} =\frac{s_k^2\big[\delta_{k+1}\xi^2 + 2s_k\xi(1-\xi) + \delta_k(1-\xi)^2\big]}{\big[s_k + \big[\delta_{k+1}+\delta_{k}-2s_k\big]\xi(1-\xi)\big]^2}
\end{align}

Inverting the spline is possible by inverting equation~(\ref{eq:rational-quadratic-spline}) and solving for the roots of the resulting quadratic equation.


\section{Methods}
\label{sec:methods}

This section describes our data preprocessing scheme (\ref{section:data}), the implementation of \spectre\ (\ref{section:model}), and our training (\ref{section:training}) and model selection (\ref{section:modelSelection}) procedures. Additionally, we describe the likeness of high-z targets to our training dataset (\ref{section:LR}) and our measurement of the red damping wing (\ref{section:dampingWing}).

\subsection{Data}
\label{section:data}

\subsubsection{Training Data}
\label{section:training_data}

\begin{figure*}
	\includegraphics[width=\textwidth]{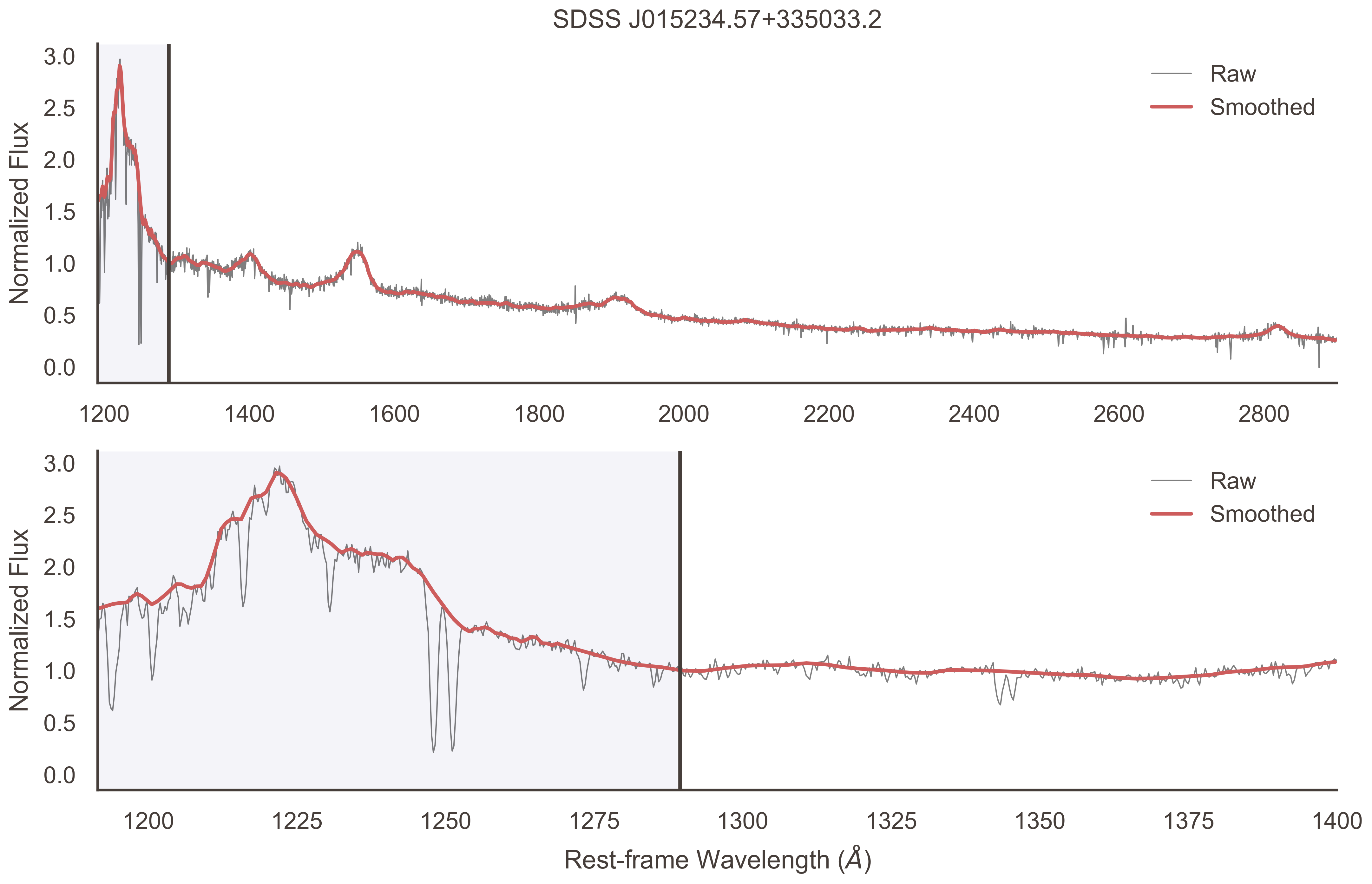}
    \caption{\textbf{Top}: an example of our preprocessing scheme. Here, we show a smoothed spectrum in red overlaid upon its raw flux counterpart in grey. The region with grey background on the left-hand side is the blue-side of the spectrum whose distribution we attempt to model conditional on the red-side of the spectrum (white background, right). \textbf{Bottom}: we zoom in on rest-frame wavelengths near Lyman-$\alpha$ to exhibit the narrow absorption features in the raw flux which have been eliminated in our preprocessing routine by identifying outliers in the difference of the raw flux and its upper envelope. All smoothed spectra are normalized to unity at $\SI{1290}{\angstrom}$, the threshold between the red and blue regions.}
    \label{fig:preprocessing_full}
\end{figure*}

We adopt the data preprocessing scheme of \cite{qsanndra}, and briefly recount it here. For a detailed description, refer to \citet{qsanndra}. A full Python implementation is available in a GitHub repository here: \href{https://github.com/DominikaDu/QSmooth}{\ttfamily github.com/DominikaDu/QSmooth}.

We select all quasar spectra from the 14th data release (DR14) of the Sloan Digital Sky Survey (SDSS) quasar catalog \citep{dr14} within the redshift range $\verb|Z_PIPE| \in [2.09,\ 2.51]$. These spectra were captured by the extended Baryon Oscillation Spectroscopic Survey (eBOSS). This redshift range was chosen to minimize our exposure to systematic uncertainties in the SDSS pipeline redshift estimates---quasars within our chosen redshift range include prominent emission features from Lyman-$\alpha$ to Mg \textsc{ii}, which strongly constrain the SDSS redshift estimates \citep{greigLya}.

We discard all spectra which are flagged as having broad absorption lines (\verb|BI_CIV| $\neq$ 0) or tenuous redshift estimates (\verb|ZWARNING| $\neq$ 0). We then discard spectra with low signal to noise ratios (\verb|SN_MEDIAN_ALL| < 7.0).

The spectra are subsequently smoothed. We begin by smoothing each spectrum with a median filter of kernel size $k=50$. Then, a peak-finding algorithm identifies any peaks of the original spectrum lying above the median-smoothed boundary. We interpolate between the peaks to create an upper envelope of the spectrum. The upper envelope is then subtracted from the original spectrum. Absorption features are readily identifiable in these residuals using a RANSAC regressor \citep{ransac} fit on the residual flux as a function of wavelength. We then interpolate between RANSAC inliers and smooth the resulting spectrum once more with a median filter of kernel size $k=20$.

After shifting each spectrum to its rest-frame using the SDSS pipeline redshift estimates, each spectrum is normalized such that its flux is unity at $\lambda_{\text{rest}} = \SI{1290}{\angstrom}$. To further clean our dataset, we eliminate any spectra whose normalized flux falls below $0.5$ blueward of $\SI{1280}{\angstrom}$ or below $0.1$ redward of $\SI{1280}{\angstrom}$. These cuts eliminate spectra with blue-side absorption which may contaminate measurements of the damping wing and spectra with a low signal-to-noise ratio on their red-side, respectively. Each spectrum was then interpolated to a fixed grid of 3,861 wavelengths between \SI{1191.5}{\angstrom} and \SI{2900}{\angstrom} spaced uniformly in log space.

The dataset was then filtered using a random forest to cull any remaining spectra with strong absorption features on the blue-side. After standardizing each spectrum such that the flux in each wavelength bin is z-score normalized, we perform an independent principal component analysis (PCA) on the blue and red-sides, then select the PCA coefficients which together explain 99\% of the dataset variance on each side. We train a random forest regressor to predict the blue-side coefficients given the red-side coefficients using 10-fold cross-validation. Outlying spectra (with strong absorption features) preferentially occupy a tail of the reconstruction error distribution which we then select upon to eliminate data points beyond three standard deviations of the mean reconstruction error. 

Our final dataset contains 13,703 quasar continua with high signal-to-noise ratios and low contamination from absorption features near Lyman-$\alpha$. Our blue- and red-side continua are composed of flux values across 345 and 3516 wavelength bins, respectively. The dataset is identical to the dataset used in \citet{qsanndra}. We divide the dataset into training, validation and testing partitions using 90/5/5 percent of the data, respectively. The partitions are chosen randomly. An example spectrum and its associated smoothed continuum approximation is shown in Fig. \ref{fig:preprocessing_full}. 

During training and inference, all input spectra were pixel-wise z-score normalized such that the model operated directly on flux z-scores calculated individually in each wavelength bin. Blue-side continua predictions were then inverse transformed before all subsequent analyses.

\subsubsection{High-z Data}
\label{section:highz_data}

Our inference targets are two high-z quasars: ULAS J1120+0641 $(z = 7.09)$ \citep{J1120} observed by VLT/FORS and Gemini/GNIRS and ULAS J1342+0928 $(z = 7.54)$ \citep{J1342} observed by Magellan/FIRE and Gemini/GNIRS.

The spectra of ULAS J1120+0641 ($z=7.09$) and ULAS J1342+0928 ($z=7.54$) contain regions of poor signal-to-noise or missing data which must be imputed in order to predict their blue-side continua. To accomplish this, we again adopt the methods from \citet{qsanndra}. We trained two fully connected feed-forward neural networks to fill in missing spectral features, one to be applied on each high-z spectrum individually. Each neural network had three hidden layers of width (55, 20, 11) neurons with exponential linear unit (ELU) activation functions. The networks were trained for 400 epochs with a batch size of 800.

For ULAS J1120+0641, the neural network inputed fluxes from $1660-\SI{1800}{\angstrom}$ and  $2200-\SI{2450}{\angstrom}$. For ULAS J1342+0928, missing data was imputed in the regions between $1570-\SI{1700}{\angstrom}$ and  $2100-\SI{2230}{\angstrom}$. After reconstructing the red-side spectra, we apply the same pre-processing pipeline as used on the moderate redshift quasars described above in Sec.~\ref{section:training_data}.

\subsection{Model}
\label{section:model}

The \spectre\ architecture is based off of \citet{neuralSplines}'s original implementation of rational quadratic neural spline flows. 

Our network employs 10 layers of spline coupling transforms, each parameterized by a residual network conditioner with 256 hidden units. The conditioner uses batch normalization and is regularized via dropout with $p=0.3$. Each spline is composed of 5 bins in the region $x,\ y \in [-10,\ 10]$ i.e. $B=10$ though we note little difference for various choices of $B$ so long as it is greater than ${\sim}3$ (but note this depends on the normalization of your data).

An encoder network is used to extract relevant information from the redward spectrum during training and inference. The encoder is a fully-connected network with 4 layers of 128 hidden units. The dimensionality reduction offered by the encoder allowed us to build very deep conditional flows without reaching our GPU's memory limit.

In summary, \spectre\ produces plausible blue-side continua by transforming random Gaussian samples through a series of ten coupling transforms, each conditioned on the red-side emission. The coupling layers sequentially contort Gaussian-distributed samples to samples from the distribution over blue-side continua. The output of our model is a z-score normalized spectrum which is ultimately re-scaled to produce a candidate sample.

A full PyTorch \citep{pytorch} implementation of \spectre\ is available on GitHub (\href{http://www.github.com/davidreiman/spectre}{\ttfamily github.com/davidreiman/spectre}).

\subsection{Training}
\label{section:training}

\spectre\ was trained on an NVIDIA V100 with a batch size of 32 and an initial learning rate of $5\mathrm{e}{-4}$. The learning rate was cosine annealed with warm restarts\footnote{Cosine annealing is a technique to reduce the learning rate over time. Lower learning rates near the end of training allow a model to settle into minima of the error manifold. Warm restarts reinitialize the learning rate and begin a new annealing schedule. These restarts have empirically been shown to reduce the wall clock time to convergence and in some cases improve model performance.} to a minimum learning rate of $1\mathrm{e}{-7}$. The initial annealing period was 5000 batches and grew by a factor of 2 after each restart. After the second restart, the learning rate was annealed to $1\mathrm{e}{-7}$ once more, after which it remained constant. \spectre's gradient norms were also clipped such that $|\nabla_{\boldsymbol{\theta}} \mathcal{L}| = \text{min}\ (|\nabla_{\boldsymbol{\theta}} \mathcal{L}|,\ 5)$. This was enforced prior to each optimizer step and was used to stabilize training by constraining the update step size in parameter space for sizeable gradients. For all of our experiments, we used the Adam optimizer \citep{adam} with $\beta_1=0.9$ and $\beta_2=0.999$.

\begin{figure}
	\includegraphics[width=\columnwidth]{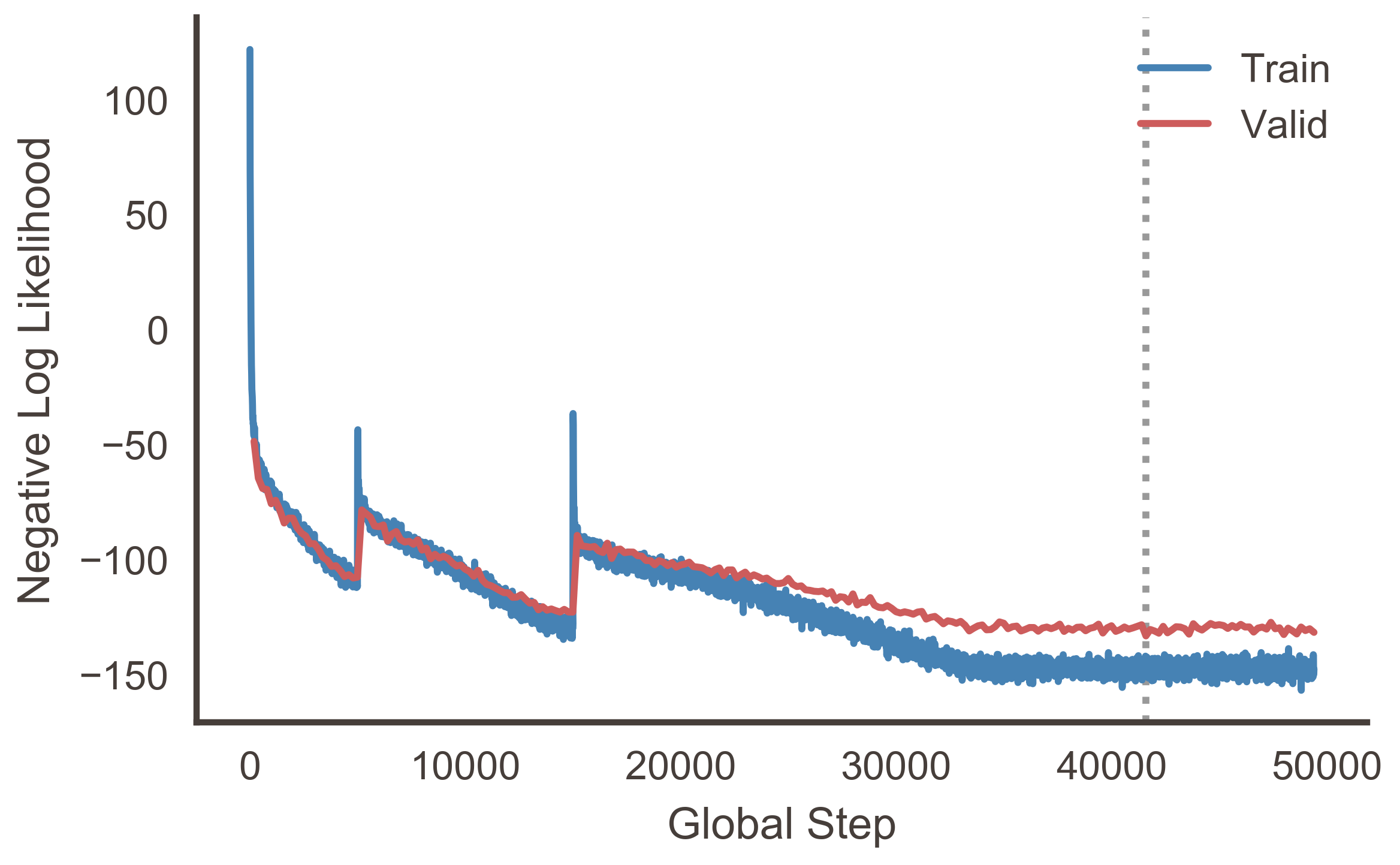}
    \caption{\spectre's training curve, showing training and validation losses (negative log likelihoods) as a function of global step, where the global step counter is incremented after each parameter update. We employ a cosine annealing learning rate schedule with warm restarts. The annealing period is $\tau=5000$ global steps and is multiplied by two after each warm restart. The dotted line marks the global step at which our model reached minimum validation error---we use the model from this step in all of our experiments.}
    \label{fig:training_curve}
\end{figure}

\subsection{Model Selection}
\label{section:modelSelection}

Our model hyperparameters were selected via an extensive grid search using a validation set. We found that small batch sizes generally yielded better generalization performance, though when the batch size was too small ($N<16$) training was often unstable and would occasionally diverge. We note that smaller models tended to perform best (likely due to the limited size of our dataset) though we explored deep conditional flows with up to one hundred coupling layers. We also found that reducing the resolution of our spectra by a factor of 3 improved the performance of our model. This was done by selecting flux values in every third wavelength bin. To verify that emission line profiles were not altered by this reduction in resolution, we compared a sample of low-resolution spectra to their unaltered counterparts and found no such issues. This downsampling cut the dimensionality of our blue- and red-side continua to 115 and 1172, respectively. We hypothesize that performance gains from this modification are due to the reduction in our data dimensionality which makes the task of modeling the density simpler for the flow. In addition, the unaltered spectra are strongly autocorrelated in such a way that downsampling does not remove a sizeable amount of information. Finally, we experimented with convolutional layers to encode red-side continua, but found they were not as effective as fully connected layers. We postulate that this is due to some underlying global features inherent in the spectra. These features could theoretically be accessed by increasing the receptive field of deep layers in the convolutional encoder, for example by adding additional layers or using dilated convolutions, but in practice we found a simple fully connected encoder worked best. A table denoting our complete model configuration is provided in Appendix~\ref{app:model_hyperparams}.

\subsection{Likeness of High-z and Moderate-z Spectra}\label{section:LR}

We explore the applicability of our primary model by using our secondary model to quantify the similarity between moderate and high redshift quasar spectra. Since normalizing flows model the likelihood of data explicitly, it naively makes sense to use these likelihoods as a measure of how \textit{in-distribution} a given continuum lies. As pointed out in \citet{gen-what} and \citet{waic}, however, this can often fail. In \citet{OODLRs}, the authors show this is an expected failure mode when semantic/informative features are sparse compared to the dimensionality of the data. Our dataset falls in this regime since it is possible to reconstruct spectra with percent-level error by using only tens of PCA components to recreate fluxes across thousands of wavelength bins \citep{daviesPCA}. 

To avoid the spurious outlier detection performance of pure likelihoods, we employ the methods of \citet{OODLRs} to quantify the notion of a spectrum being in-distribution with the likelihood ratio:

\begin{equation}
    \ell(\mathbf{x}) = \frac{\pin(\mathbf{x})}{\pout(\mathbf{x})}
\end{equation}
where $\pin$ is the likelihood of datum $\mathbf{x}$ given by a model trained on in-distribution data and $\pout$ is the likelihood of $\mathbf{x}$ given by a model trained on out-of-distribution (OOD) data.

It is instructive to consider the case where semantic and background features are independently generated. In this scenario, one can split the likelihood of a sample, $\mathbf{x}$, into $p(\mathbf{x}) = p(\mathbf{x}_S)p(\mathbf{x}_B)$, where $\mathbf{x}_S$ are semantic features and $\mathbf{x}_B$ are background features. When semantic features are sparse, the likelihood $p(\mathbf{x})$ is dominated by the uninformative background. If both $\pin$ and $\pout$ give approximately the same density estimate for the background, the full likelihood ratio reduces to a likelihood ratio of semantic information. In the dependent case where background and semantic features are not independently generated, background  dependence cannot be eliminated, but the likelihood ratio can still be approximated as:

\begin{equation}
    \ell(\mathbf{x}) = \frac{\pin(\mathbf{x}_S | \mathbf{x}_B)\pin(\mathbf{x}_B) }{\pout(\mathbf{x}_S | \mathbf{x}_B)\pout(\mathbf{x}_B) } \approx \frac{\pin(\mathbf{x}_S | \mathbf{x}_B) }{\pout(\mathbf{x}_S | \mathbf{x}_B)}
\end{equation}
In practice, one does not always have access to an OOD dataset. With the correct noise model, however, perturbations can be added to the in-distribution dataset which preserve population-level background statistics, but corrupt in-distribution features.

For our application, the in-distribution data are the cleaned spectra described in Sec.~\ref{section:training_data}. We tested different noise models to create the out-of-distribution data and we found the dataset containing the unprocessed flux measurements to give the most stringent OOD limits for both ULAS J1120+0641 and ULAS J1342+0928. Results and noise models are summarized in Appendix~\ref{app:LR}. 

Because this is a density estimation exercise, we choose to use an autoregressive transform for this rational quadratic neural spline flow. We train two such flows --- one on the in-distribution dataset and one on the out-of-distribution dataset --- to map the red-side spectra onto a multivariate Gaussian which can be evaluated to obtain $\pin$ and $\pout$. These models contain the same hyperparameters listed in Table~\ref{tab:hyperparameters} aside from the number of encoder layers since there is no conditional information for this task.

\begin{figure}
	\includegraphics[width=\columnwidth]{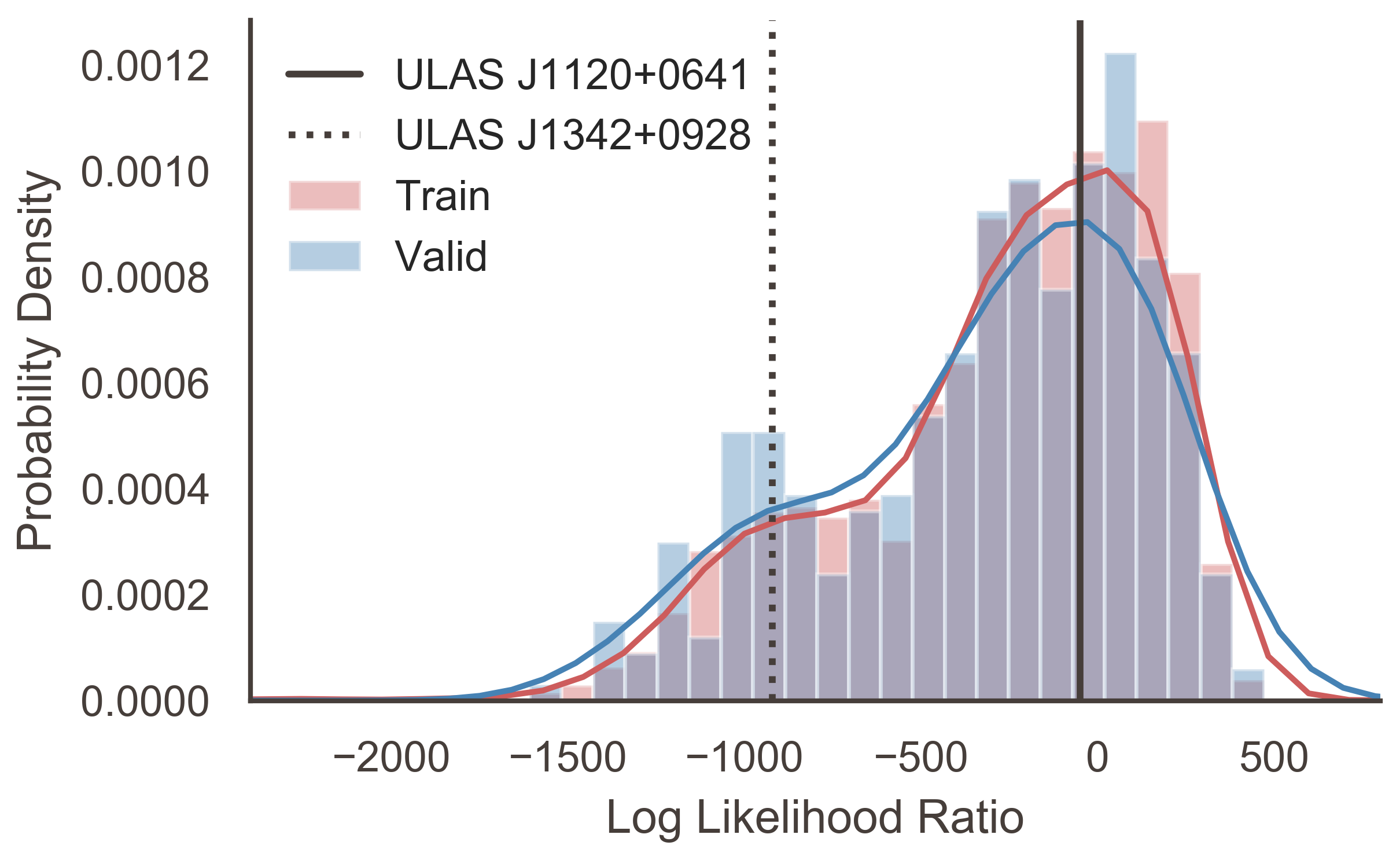}
    \caption{We evaluate the likelihood ratios of our training/validation sets and high-z spectra as calculated by two autoregressive rational quadratic neural spline flows. The most stringent out-of-distribution bounds result when one flow is trained on smoothed continua and the other flow is trained on raw flux measurements. We find  ULAS J1120+0641 and ULAS J1342+0928 lie in the 61.1 and 10.4 percentiles of our training set, respectively. This implies that both high-z quasars share significant likeness to our training distribution and are valid inputs to our primary model.}
    \label{fig:LRs}
\end{figure}

In Fig.~\ref{fig:LRs} we show the likelihood ratios of training and validation sets along with high-z spectra as calculated by the two flows. Training and validation sets overlap showing the models have not overfit on the training set. As there are no guarantees for OOD detection using this method, we treat a sample's likelihood-ratio percentile as an upper bound on how in-distribution a sample lies. The likelihood ratio of ULAS J1120+0641 $(z=7.09)$ falls in the 61.1 percentile of our training set which means it is well represented by our training data. ULAS J1342+0928 $(z=7.54)$ is in the 10.4 percentile of our training set which, although not an outlier, is less typical of a moderate redshift continuum. 

This hierarchy holds across various methods in the literature. In \citet{daviesPCA} the likelihood of 10 red-side PCA coefficients from a Gaussian mixture model is used to give 15 and 1.5 percentiles to ULAS J1120+0641 and ULAS J1342+0928, respectively. In \citet{qsanndra}, an autoencoder is trained to reconstruct 63 red-side coefficients. The reconstruction error of the red-side coefficients then gives a quantifiable measure of how well represented the high redshift continua are by the training set. With this method, they assign 52 and 1 percentiles to  ULAS J1120+0641 and ULAS J1342+0928, respectively. 

\subsection{Measurement of the Damping Wing}
\label{section:dampingWing}

\begin{figure}
	\includegraphics[width=\columnwidth]{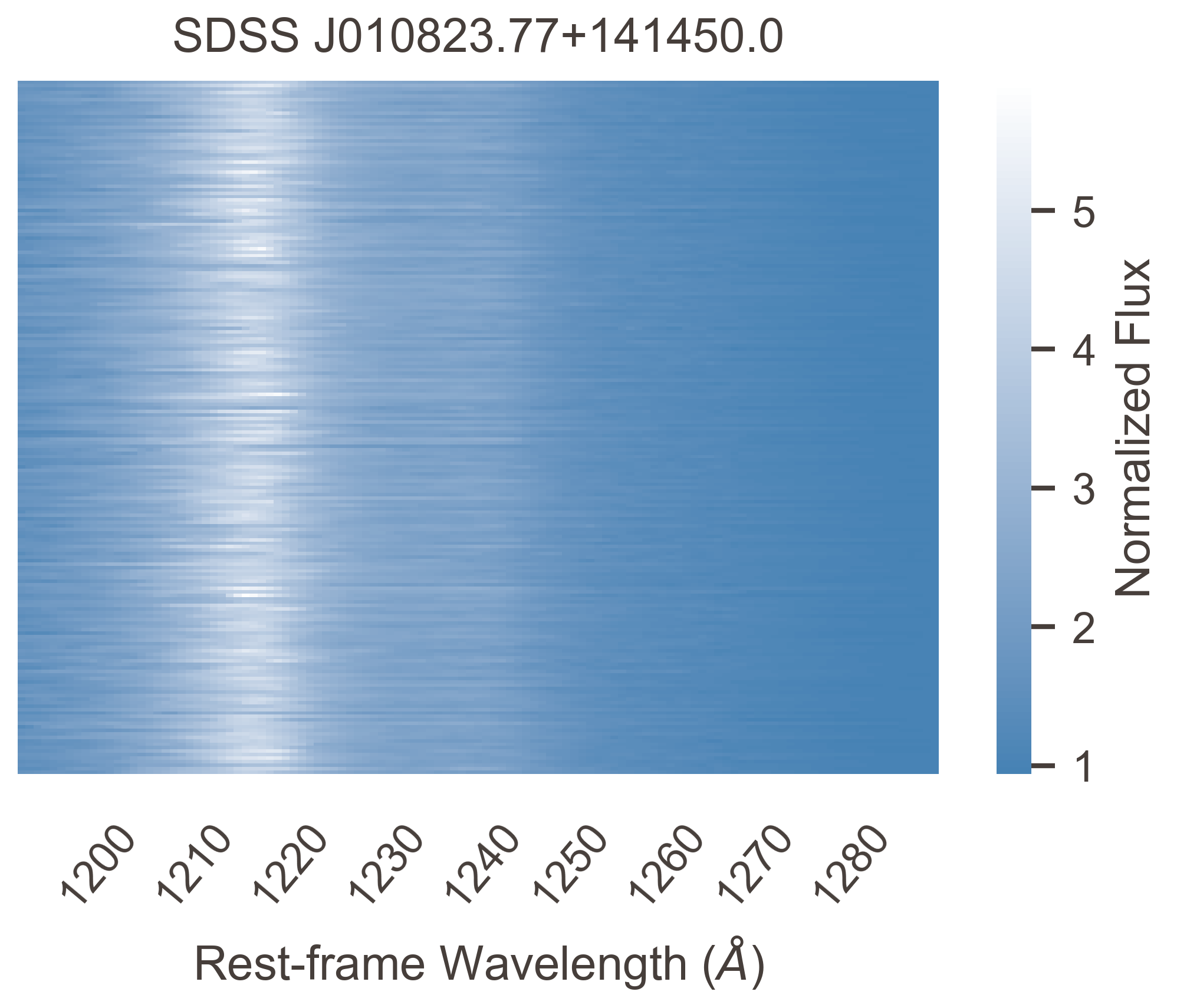}
    \caption{A selection of 200 blue-side continuum predictions from \spectre\ for a randomly chosen quasar. Each row corresponds to a single blue-side sample from our model, color-coded to designate its normalized flux at each wavelength. Qualitatively, the model's predictions are consistent with a \lya\ peak near $\SI{1215.67}{\angstrom}$ and a N \textsc{v} emission near $\SI{1240.81}{\angstrom}$ .} 
    \label{fig:sunset}
\end{figure}
\begin{figure*}
	\includegraphics[width=\textwidth]{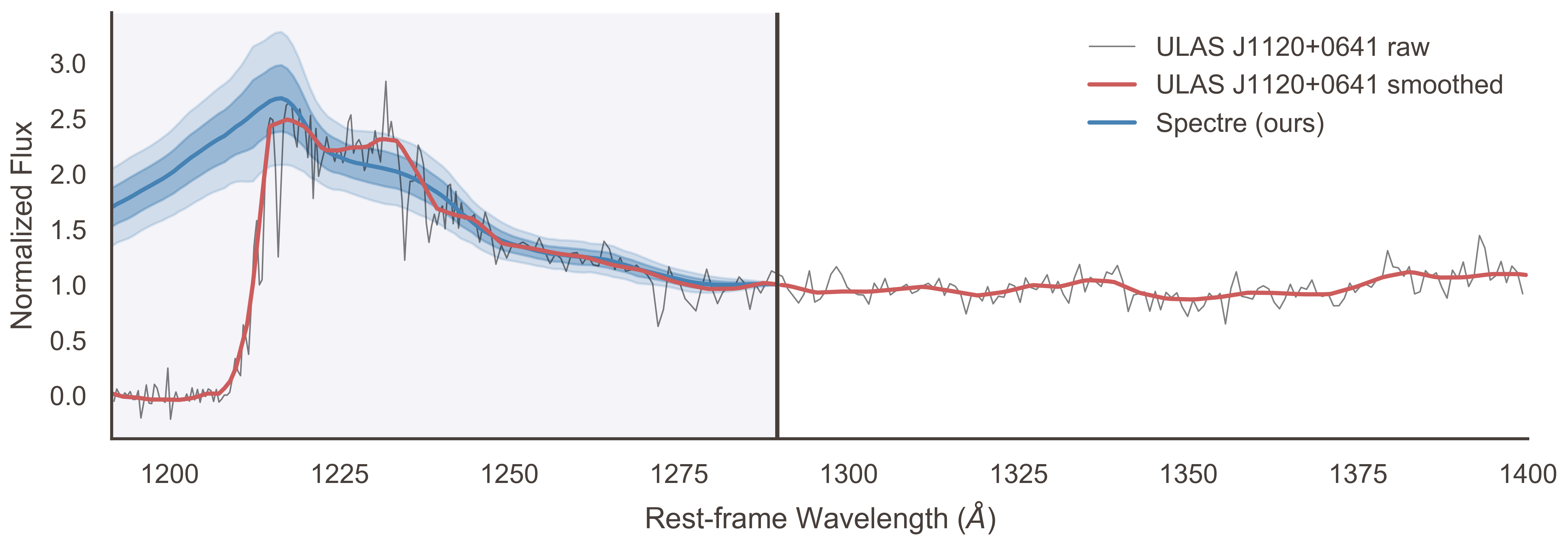}
	\includegraphics[width=\columnwidth]{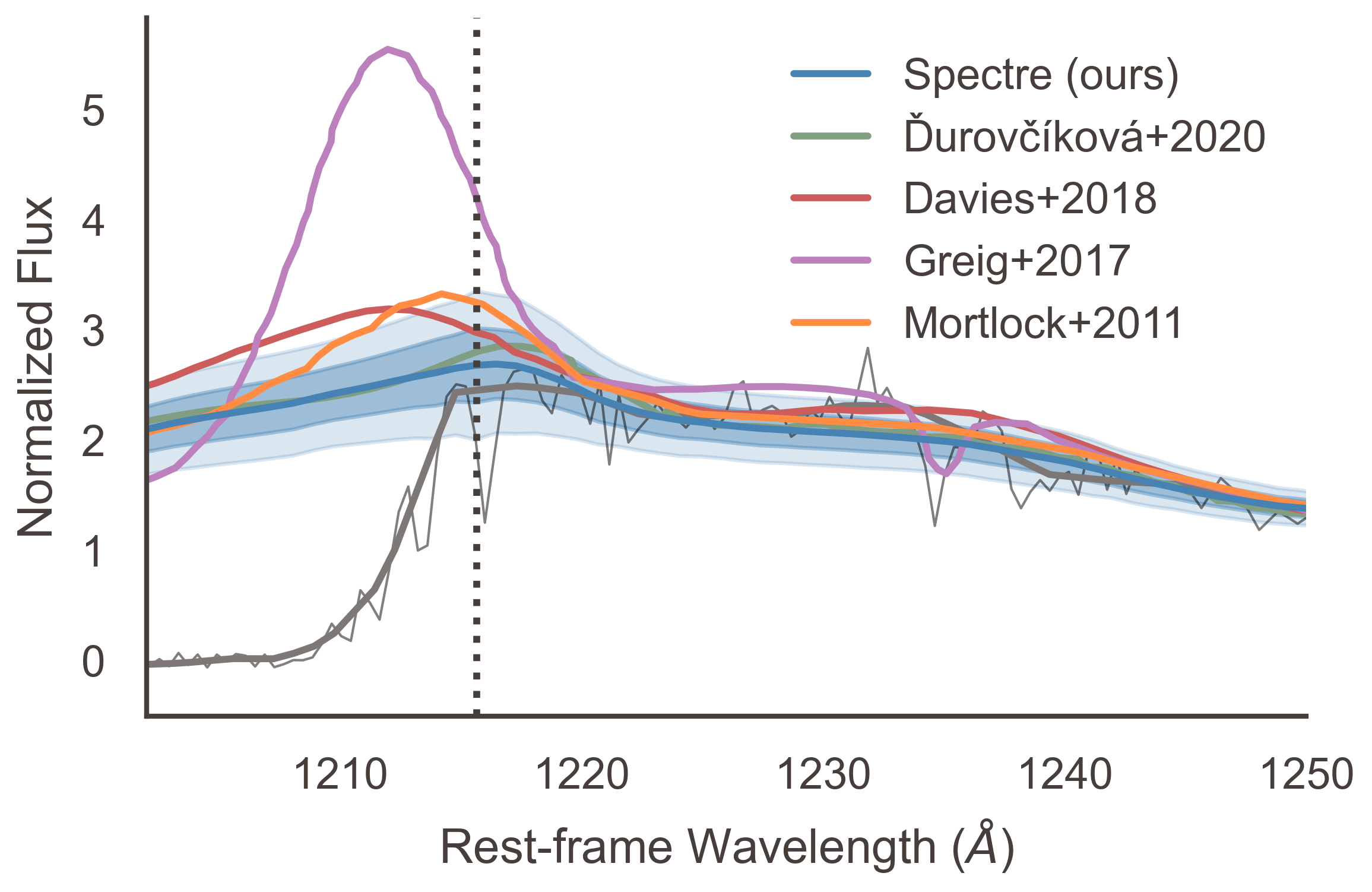}
	\includegraphics[width=\columnwidth]{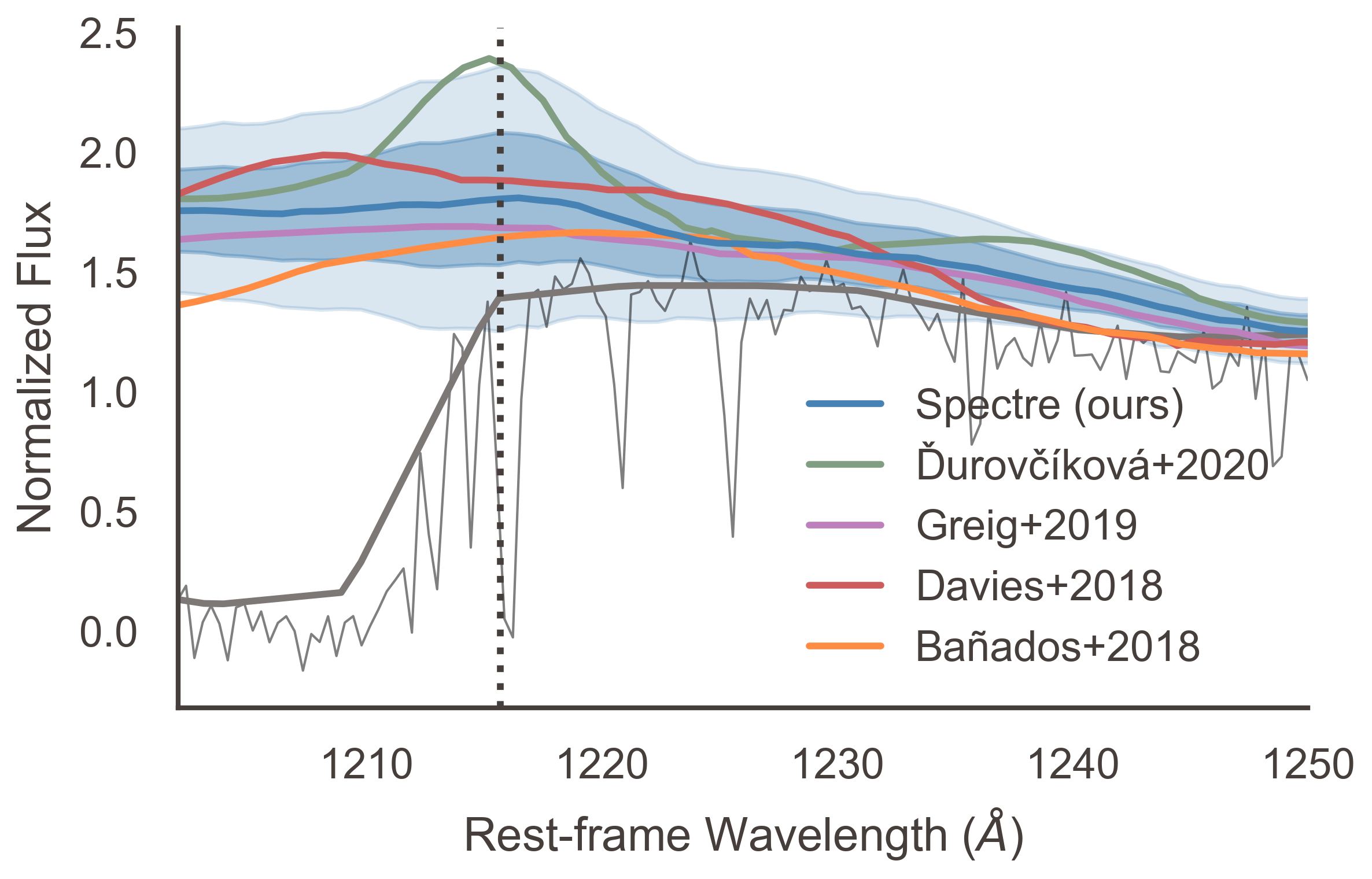}
	\includegraphics[width=\textwidth]{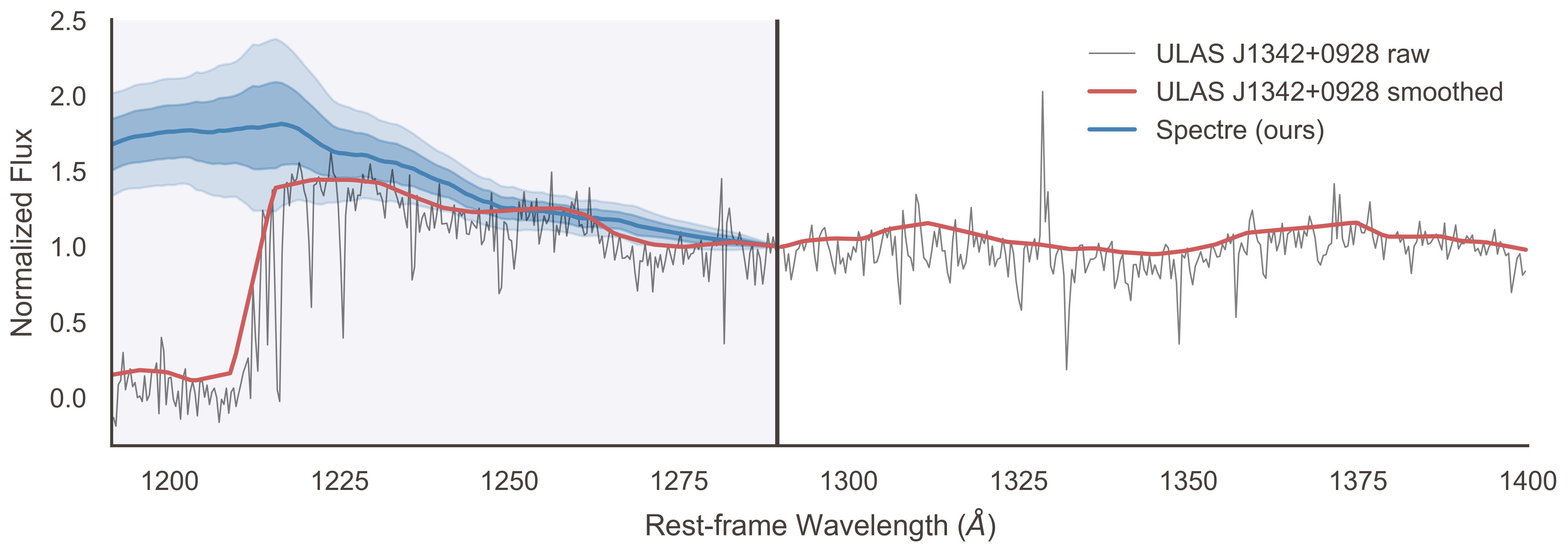}
    \caption{Blue-side continua predictions on two high redshift quasars, ULAS J1120+0641 and ULAS J1342+0928. Each solid blue line is the mean of one thousand samples from \spectre. The blue and light blue bands reflect one and two standard deviation bands at each wavelength, respectively. \textbf{Top}: Our mean prediction with two sigma uncertainty bands overlaid on top of continuum and raw flux measurements of ULAS J1120+0641. \textbf{Bottom}: Our mean prediction with two sigma uncertainty bands overlaid on top of continuum and raw flux measurements of ULAS J1342+0928. \textbf{Middle Left}: A comparison of predictions from various authors on  ULAS J1120+0641.  \textbf{Middle Right}:  A comparison of predictions from various authors on ULAS J1342+0928. }
    \label{fig:all_highz}
\end{figure*}

To estimate the neutral fraction of hydrogen near the epoch of reionization, we measure the damping wing of the Gunn-Peterson trough in two high-redshift quasars: ULAS J1120+0641 at $z=7.09$ \citep{J1120} and ULAS J1342+0928 at $z=7.54$ \citep{J1342}. Measurement of the damping wing requires knowledge of the intrinsic emission of each quasar, which \spectre\ provides using conditional information from the redward spectrum. We proceed by assuming that the IGM is uniformly neutral from $z_n$ to the blueward edge of the quasar near-zone, $z_\text{nz} = (1 + z_s)(\lambda_\text{nz}/\lambda_\alpha) - 1 $ where $z_s$ is the redshift of the source. For both ULAS J1120+0641 and ULAS J1342+0928 we use $\lambda_\text{nz} = \SI{1210}{\angstrom}$ and set $z_n = 6$.

We model the red damping wing using the analytical model of \citet{dampingWing}:

\begin{equation}
\label{eqn:damping_wing}
    \tau(\Delta\lambda) = \frac{\tau_{\text{GP}} R_\alpha}{\pi}(1+\delta)^{3/2} \int_{x_1}^{x_2} \frac{x^{9/2}}{(1-x)^2} dx
\end{equation}
where $\delta = \Delta\lambda/[\lambda_{\text{nz}}(1 + z_{\text{nz}})]$ and $\Delta\lambda = \lambda - \lambda_{\text{nz}} (1 + z_\text{nz})$ is the wavelength offset (in the observed frame) from the Lyman-$\alpha$ transition at the edge of the near-zone. The bounds of the integral are given by $x_1 = (1 + z_n)/[(1 + z_\text{nz})(1+\delta)]$ and $x_2 = (1+\delta)^{-1}$. The constant\footnote{$R_\alpha = \Lambda/(4\pi\nu_\alpha)$ with $\Lambda$ the decay constant of the Lyman-$\alpha$ resonance and $\nu_\alpha$ the frequency of the Lyman-$\alpha$ line---see \citet{dampingWing}.} $R_\alpha=2.02\times 10^{-8}$, and the Gunn-Peterson optical depth of neutral hydrogen, $\tau_{\text{GP}}$, is given by \citet{fanOpticalDepth} as follows:

\begin{equation}
    \tau_{\text{GP}}(z) = 1.8\times 10^5 h^{-1} \Omega_m^{-1/2}\Big(\frac{\Omega_bh^2}{0.02}\Big)\Big(\frac{1+z}{7}\Big)^{3/2}\bar{x}_{\text{HI}}
\end{equation}

The integral in Eqn.~\ref{eqn:damping_wing} is solvable analytically and its solution is provided in \citet{dampingWing} as:

\begin{equation}
    I(x) = \frac{x^{9/2}}{1-x} + \frac{9}{7}x^{7/2} + \frac{9}{5}x^{5/2} + 3x^{3/2} + 9x^{1/2} - \frac{9}{2}\log\frac{1 + x^{1/2}}{1-x^{1/2}}
\end{equation}

We adopt the Planck 2018 cosmological parameters \citep{planck2018} of $h = 0.6766 \pm 0.0042$, $\Omega_m = 0.3111 \pm 0.0056$ and $\Omega_b h^2 = 0.02242\pm 0.00014$. To determine the end of the proximity zone, we employ a common heuristic in the literature: the edge of the proximity zone is where the smoothed spectrum equals one tenth of its magnitude at at Lyman-$\alpha$. For both high-z quasars, this method suggests a blueward edge of ${\sim}\SI{1210}{\angstrom}$.

We aim to fit the damping wing model to the observed damping wing where the volume-averaged neutral fraction of hydrogen, $\bar{x}_{\text{HI}}$, is our only free parameter. Equipped with our predictions of the intrinsic quasar emission near Lyman-$\alpha$, $F_{\text{int}}$, we measure the observed damping wing by computing the optical depth as a function of wavelength in the range $\lambda_{\text{rest}} \in [\SI{1210}{\angstrom}, \SI{1250}{\angstrom}].$

\begin{equation}
    \tau_{\text{obs}} = -\ln\Big({\frac{F_{\text{obs}}}{F_{\text{int}}}}\Big)
\end{equation}
where $F_{\text{obs}}$ is the observed flux of the quasar for which we use the smoothed spectrum of the quasar. Note that each blue-side continua prediction sampled from \spectre\ provides a separate estimate of $F_{\text{int}}$, and thereby a new measurement of $\tau_{\text{obs}}$ and the neutral fraction $\bar{x}_{\text{HI}}$. By Monte Carlo sampling plausible continua, we can very easily estimate the distribution over the neutral fraction.

To estimate the neutral fraction itself, we assume that the value of $\tau$ in each wavelength bin is distributed according to a Gaussian distribution such that:

\begin{figure*}
    \centering
    \begin{subfigure}{\columnwidth}
      \centering
    	\includegraphics[width=\columnwidth]{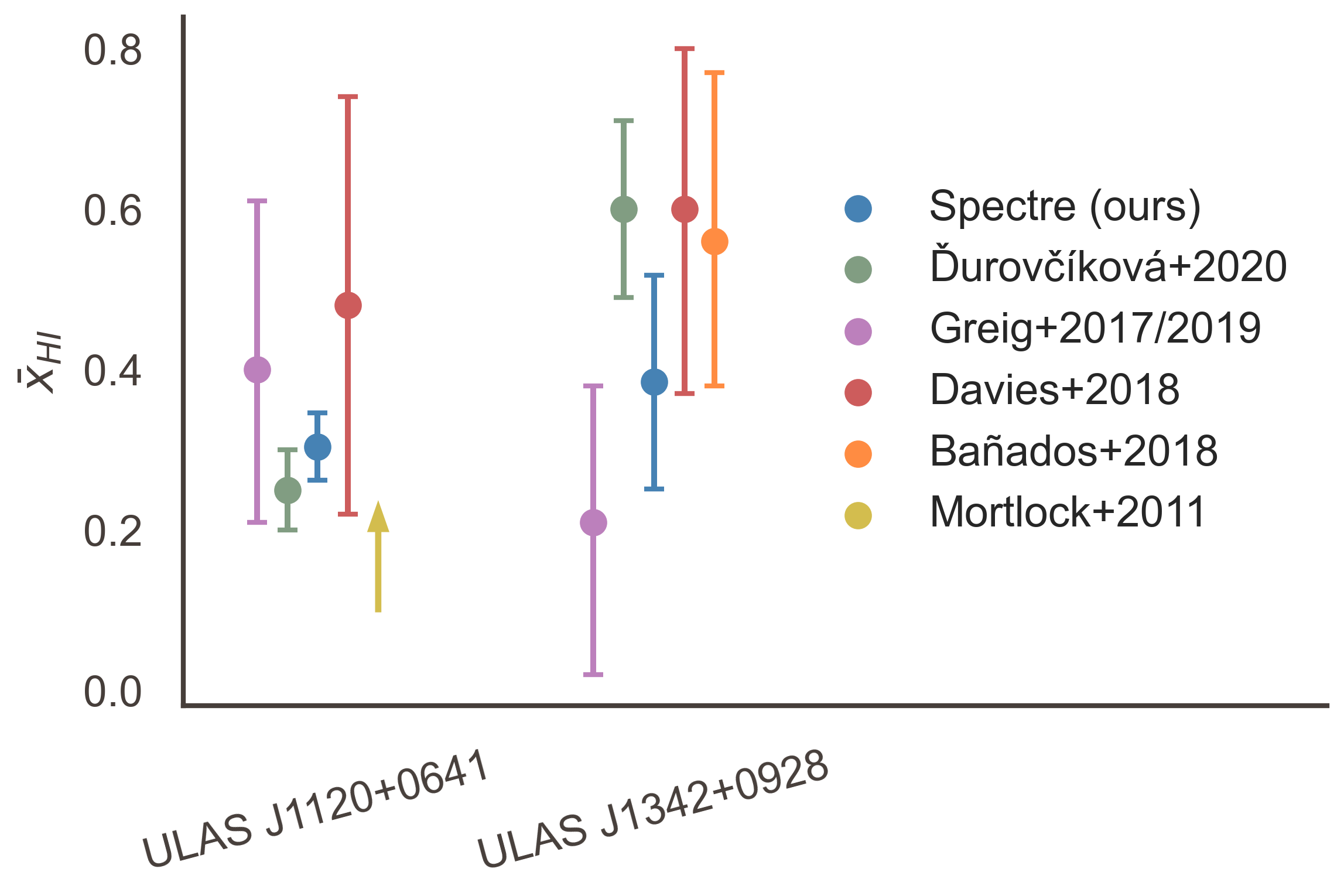}
    \end{subfigure}%
    \begin{subfigure}{\columnwidth}
      \centering
    	\includegraphics[width=\columnwidth]{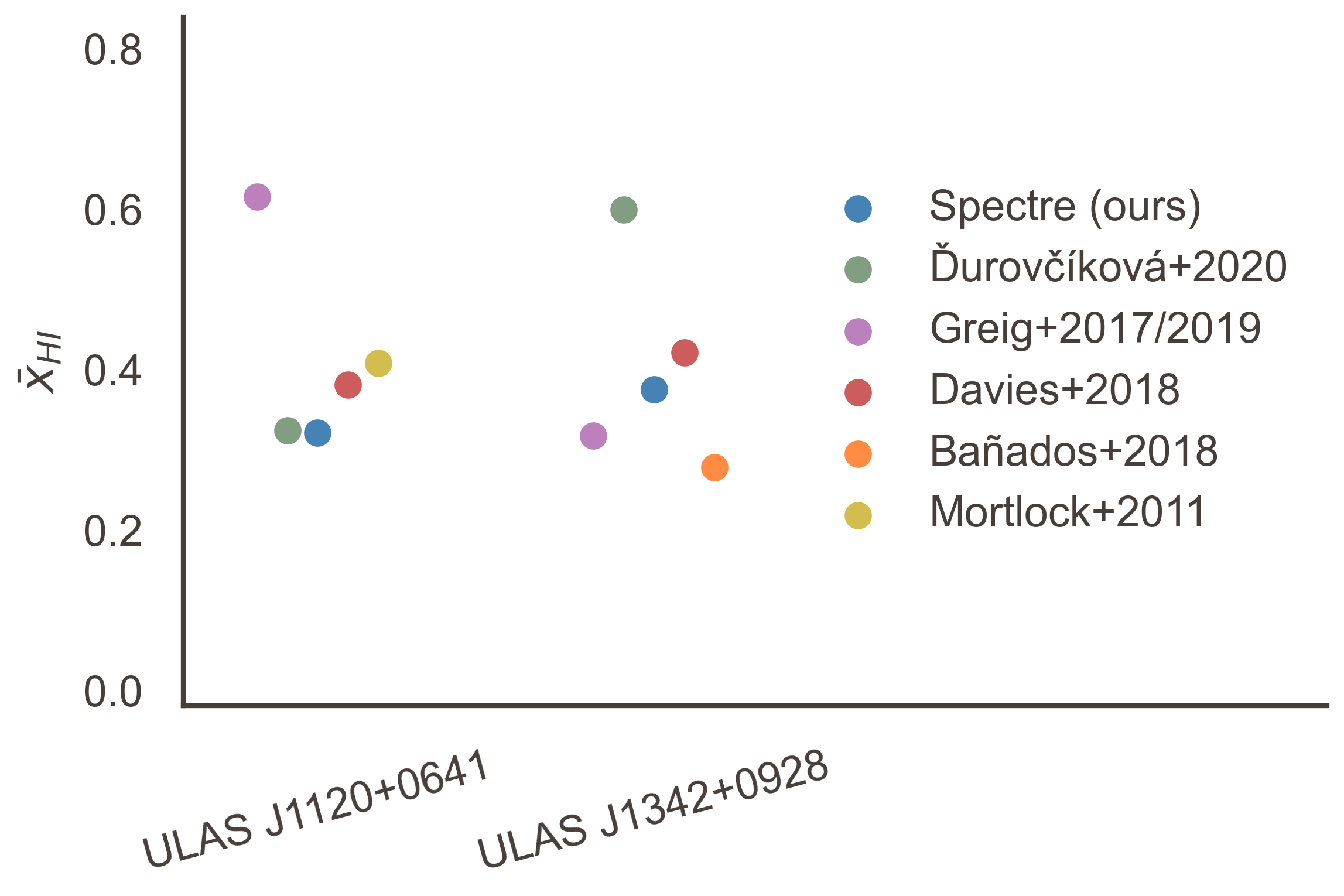}
    \end{subfigure}
    \caption{A comparison of our estimates of the volume-averaged neutral fraction of hydrogen for ULAS J1120+0641 ($z=7.09$) and ULAS J1342+0928 ($z=7.54$). \textbf{Left}: Reported results from the literature. These make use of different damping wing models (and some employ full hydrodynamical models of the IGM) which complicates direct comparison. \textbf{Right}: Neutral fractions for ULAS J1120+0641 and ULAS J1342+0928 computed from the mean intrinsic continua prediction of all previous approaches and a single damping wing model \citep{dampingWing}. Error bars are omitted since only mean continuum predictions are used.}
    \label{fig:combined_neutral_fraction}
\end{figure*}

\begin{figure}
	\includegraphics[width=\columnwidth]{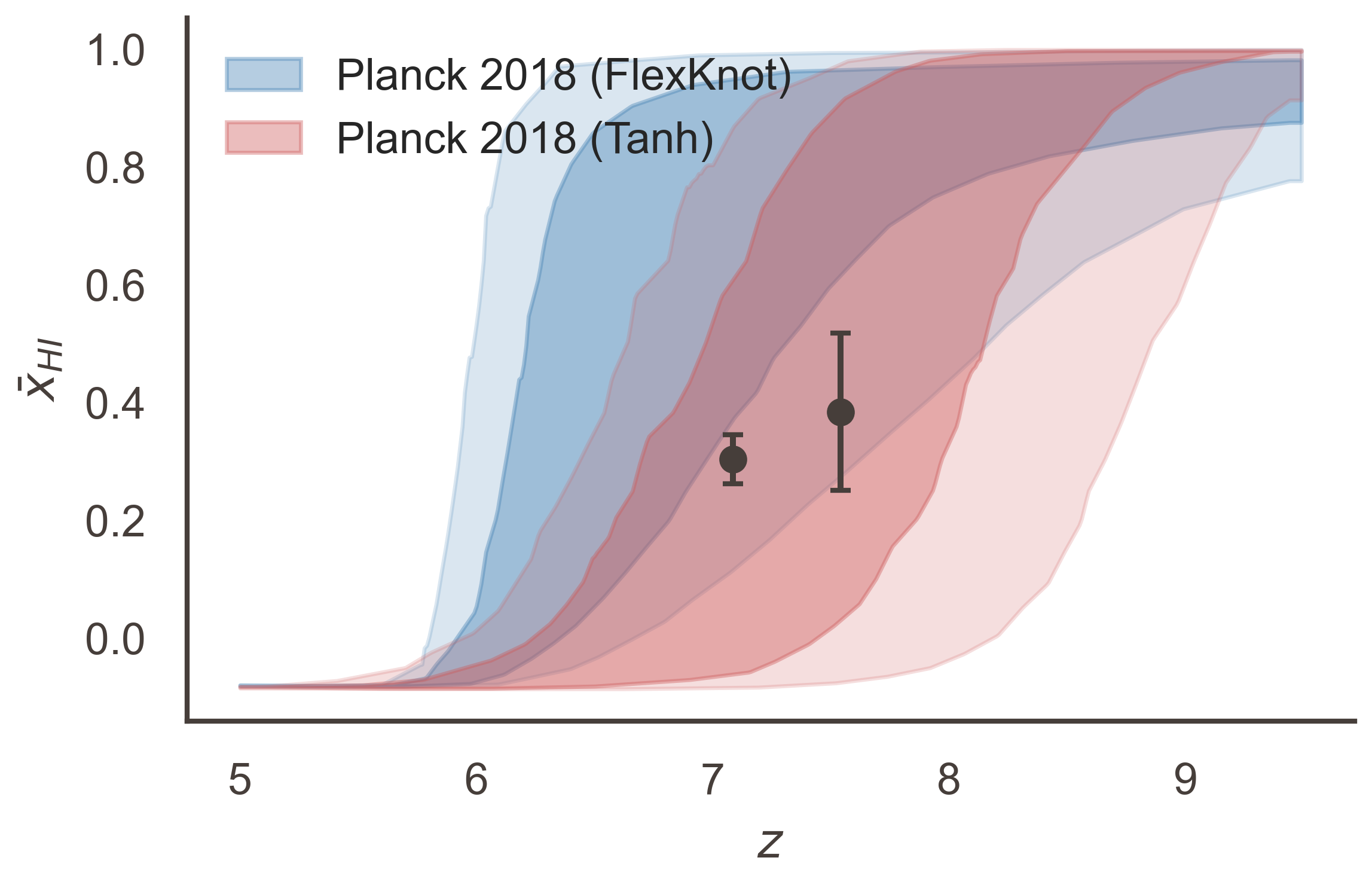}
    \caption{A comparison of our estimates of the volume-averaged neutral fraction of hydrogen to the Planck constraints \citep{planck2018}. Planck employs two models for their reionization constraints: the \textit{Tanh} model which assumes a smooth transition from a neutral to ionized universe based on a hyperbolic tangent function and the \textit{FlexKnot} model which can flexibly model any reionization history based on a piecewise spline with a fixed number of knots (though the final result is marginalized over this hyperparameter). \spectre's predictions are well within the 1-sigma confidence interval for Planck's Tanh constraints and within the 2-sigma confidence interval for the FlexKnot constraints.}
    \label{fig:planck_constraints}
\end{figure}

\begin{figure}
	\includegraphics[width=\columnwidth]{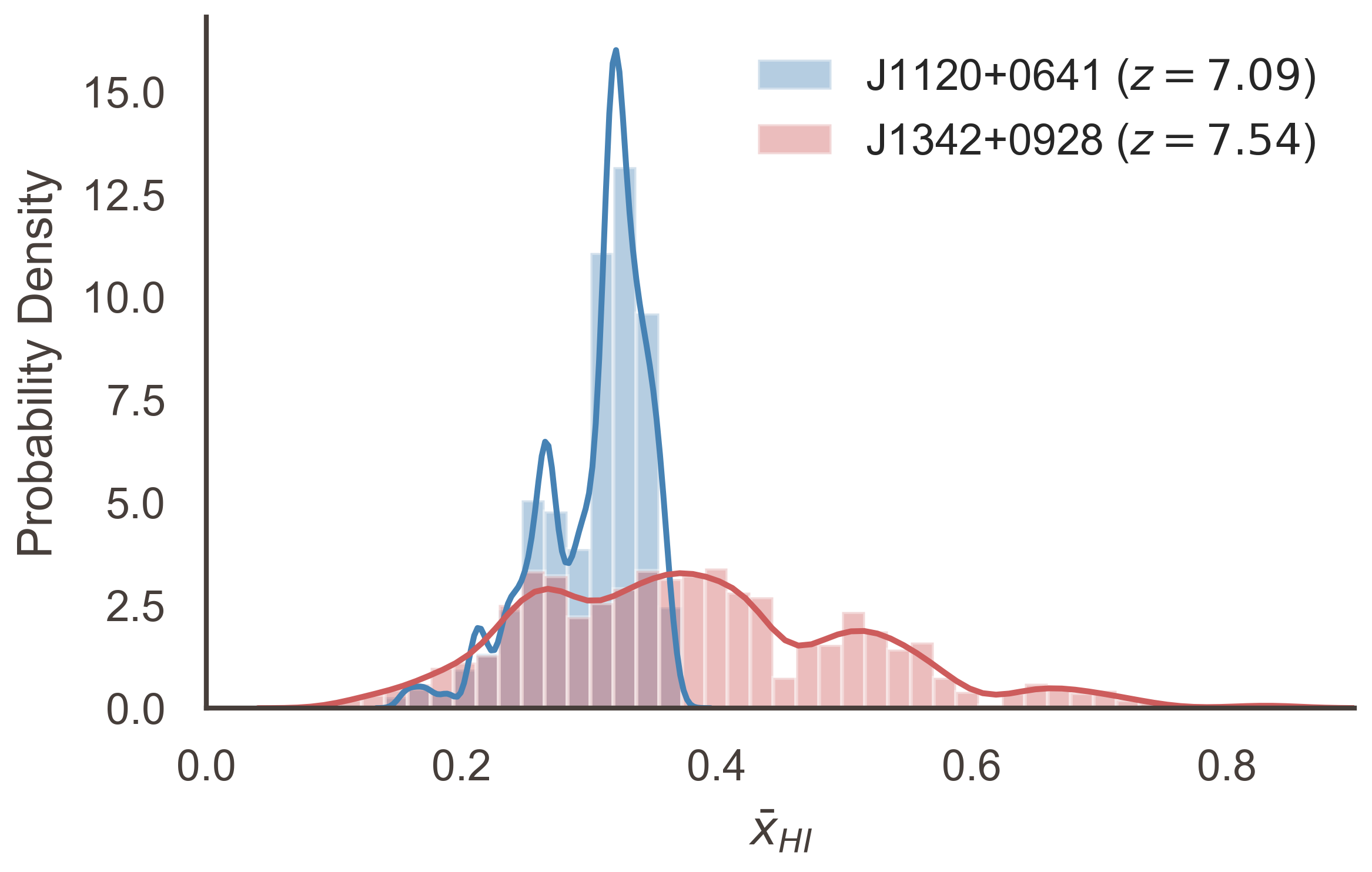}
    \caption{Histograms and kernel density estimates depicting the spread in neutral hydrogen fraction predictions over 10000 samples from \spectre. From observations of ULAS J1120+0641 we infer $\bar{x}_{\text{HI}} = \xlow$ while for ULAS J1342+0928 we infer $\bar{x}_{\text{HI}} = \xhi$.}
    \label{fig:neutral_fraction}
\end{figure}

\begin{equation}
    \tau_{\text{obs}}(\Delta\lambda) \sim \mathcal{N}(\tau_{\text{GP}} R_\alpha/\pi (\Delta\lambda/\lambda)^{-1},\ \sigma_\tau^2)
\end{equation}
and we perform maximum likelihood inference to estimate the parameter $\bar{x}_{\text{HI}}$ which is hidden inside of $\tau_{\text{GP}}$. This amounts to a non-linear least squares problem with the additional constraint that $0 \leq \bar{x}_{\text{HI}} \leq 1$. Such problems are easily solved with readily available optimization routines in Python, or a simple grid search over the open unit interval.

Uncertainty in cosmological parameters, the redshift of the source, and our estimate of the intrinsic continua all introduce error into our model. In addition, we've made simplifying assumptions: (i) the quasar's proximity zone is entirely ionized, and (ii) the IGM is uniformly dense and neutral beyond the proximity zone. In reality, the proximity zone contains residual neutral hydrogen and the IGM beyond the proximity zone is patchy and uneven, attributable to the growing ionization bubbles surrounding other luminous sources along the line-of-sight.

To estimate the effect of the uncertainty in each of the above factors to our estimates of the neutral fraction, we use a Monte Carlo approach. We treat the source redshift and the cosmological parameters as Gaussian distributed random variables for which the reported mean is the location of the mode and the uncertainty describes the standard deviation of the mean. We then proceed by:

\begin{enumerate}[i.]
    \item Sampling a source redshift
    \item Shifting the red-side spectrum to its rest-frame
    \item Estimating the blue-side continua
    \item Sampling a random vector of cosmological parameters
    \item Estimating the neutral fraction
\end{enumerate}

By repeated random sampling, we can empirically estimate the error propagation from each of these sources. We use a thousand samples of plausible blue-side continua from \spectre, and for each sample run ten Monte Carlo simulations by drawing random source redshifts and cosmological parameters. We then approximate the distribution over the neutral fraction of hydrogen with the resulting 10,000 estimates. As is typical in the literature, we quote the mean and standard deviation of these 10,000 samples as our prediction and uncertainty, though we note that (especially for ULAS J1120+0641) the distributions are notably non-Gaussian (see Fig.~\ref{fig:neutral_fraction}).

\section{Results}
\label{sec:results}


\subsection{Reionization History Constraints}

We display our predictions of the intrinsic continua of J1120+0641 and J1342+0928 in Fig.~\ref{fig:all_highz}. For visual comparison with other approaches, we've also included figures which compare our continua predictions to those found in the literature. For ULAS J1120+0641 our intrinsic continua prediction closely matches that of \citealt{qsanndra}, suggesting a modest Lyman-$\alpha$ emission. Of the previous approaches we've considered, we predict the weakest emission. Meanwhile, for ULAS J1342+0928 we predict a moderate Lyman-$\alpha$ emission which places our intrinsic continua prediction approximately midway between those of previous approaches. It should be noted, however, that \spectre's uncertainty is greater in its prediction of ULAS J1342+0928 and the mean continua predictions of all previous approaches are captured within our 2-sigma confidence interval though this is markedly untrue for ULAS J1120+0641.

Using the model described in Sec.~\ref{section:dampingWing}, we estimate the volume-averaged neutral fraction of hydrogen to be $\bar{x}_\text{HI}= \xlow$ for ULAS J1120+0641 ($z=7.0851 \pm 0.003$) and $\bar{x}_\text{HI}= \xhi$ for ULAS J1342+0928 ($z=7.5413 \pm 0.0007$). A comparison between the estimated volume-averaged neutral fraction for our approach and all previous approaches is provided on the left of Fig.~\ref{fig:combined_neutral_fraction}. We also display our results for the volume-averaged neutral fraction of hydrogen in the context of the Planck constraints \citep{planck2018} in Fig.~\ref{fig:planck_constraints}.  

We caution the reader to be wary of direct comparison to previous approaches in the literature. We note that each previous approach uses very different models of the damping wing. Some employ full hydrodynamical models of the IGM while others (such as ours) make simplifying assumptions. In an attempt to provide a more direct comparison, we've used the mean continuum prediction from each previous approach and computed the neutral fraction with a single damping wing model \citep{dampingWing}. The results are presented on the right of  Fig.~\ref{fig:combined_neutral_fraction} and show quite different results in some cases, especially for those that employed full hydrodynamical models of the IGM. This is expected and perhaps sheds some light on the extent to which simplifying assumptions about the state of the foreground IGM biases calculations of the neutral fraction.


\subsection{Bias and Uncertainty}

To evaluate \spectre, we measure our continuum bias and uncertainty on a randomly selected validation set from the collection of moderate redshift spectra gathered from eBOSS. These are $N \approx 650$ quasars which were not seen during training. We define the relative continuum error $\epsilon_c$ as follows:

\begin{align}
    \epsilon_c = (F_\text{model} - F_\text{truth}) / F_\text{truth}
    \label{eqn:epsilon}
\end{align}
where we assume the smoothed continuum estimate provided by our preprocessing method is representative of $F_\text{truth}$ and we take the average over all elements of our validation set. The relative bias is then $\langle \epsilon_c \rangle$ and the relative uncertainty $\sigma(\epsilon_c)$. Here it is important to note that this definition of $\epsilon_c$ differs from \citealt{qsanndra} as it omits the absolute value on the residual term.

\begin{figure}
	\includegraphics[width=\columnwidth]{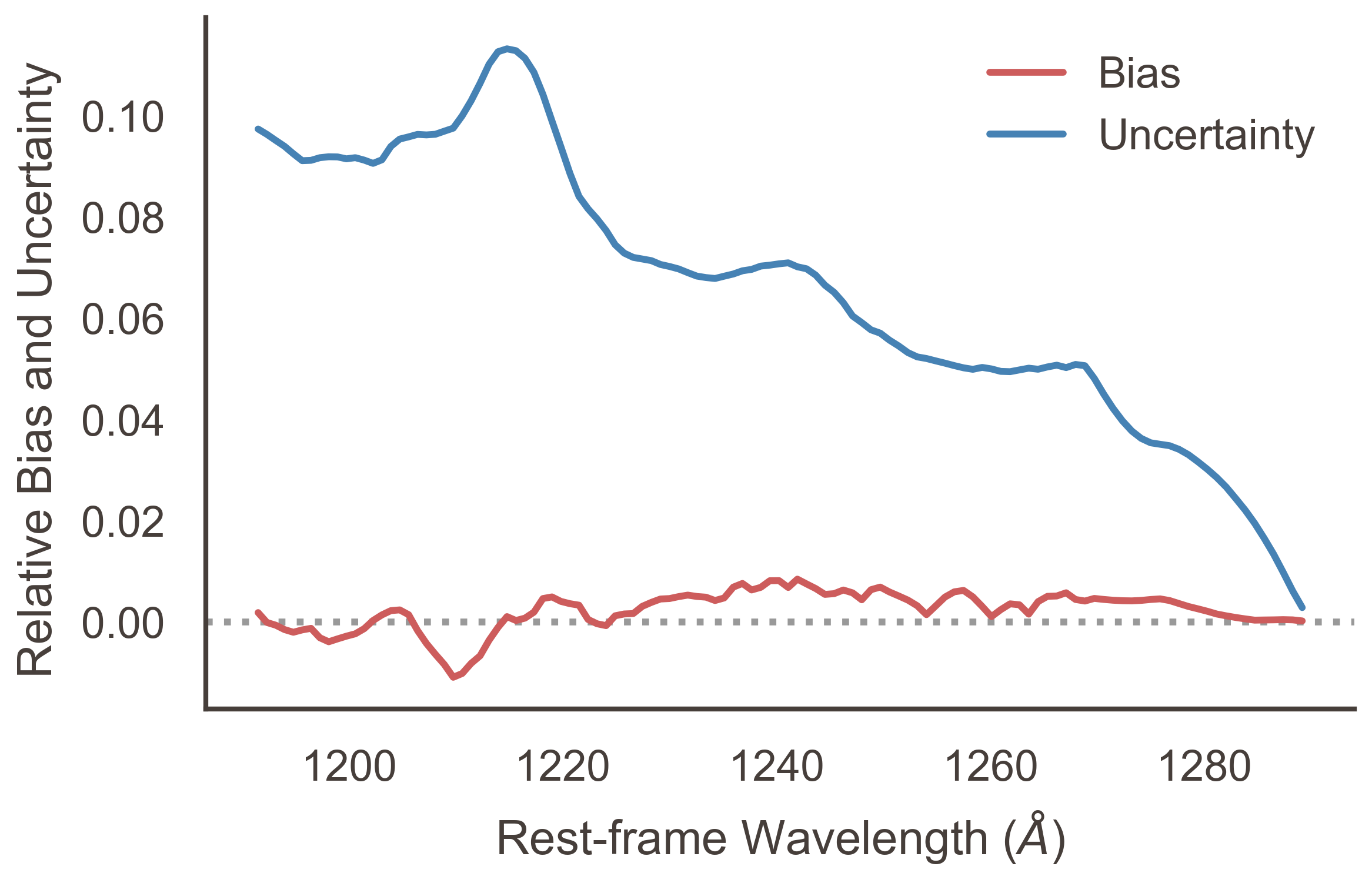}
    \caption{The relative prediction bias $\langle \epsilon_c \rangle$ and uncertainty $\sigma(\epsilon_c)$ (see Eqn.~\ref{eqn:epsilon}) as a function of blue-side wavelength averaged over our test set. Our average relative prediction uncertainty across all blue-side wavelengths is 6.63\%, though we note that this metric is highly sensitive to the preprocessing scheme and is therefore difficult to compare to other methods.}
    \label{fig:bias_uncertainty}
\end{figure}

In Fig.~\ref{fig:bias_uncertainty}, we show our relative bias and uncertainty as a function of blue-side wavelength averaged over the validation set. We maintain low relative uncertainty at all wavelengths, averaging $6.63\%$ across all blue-side wavelengths. However, we note that this metric is very difficult to compare between approaches since it is strongly dependent upon the preprocessing scheme. The relative uncertainty trends downward as the continuum approaches $\lambda_{\text{rest}} = \SI{1290}{\angstrom}$, the threshold between the blue and red-side spectrum where the extrapolation becomes trivial. Our model is largely unbiased save for a tendency to very slightly overpredict the continua redward of \lya. We average a relative bias of $0.34\%$ and are notably not strongly biased near the peak of the \lya\ emission itself.

\begin{figure}
	\includegraphics[width=\columnwidth]{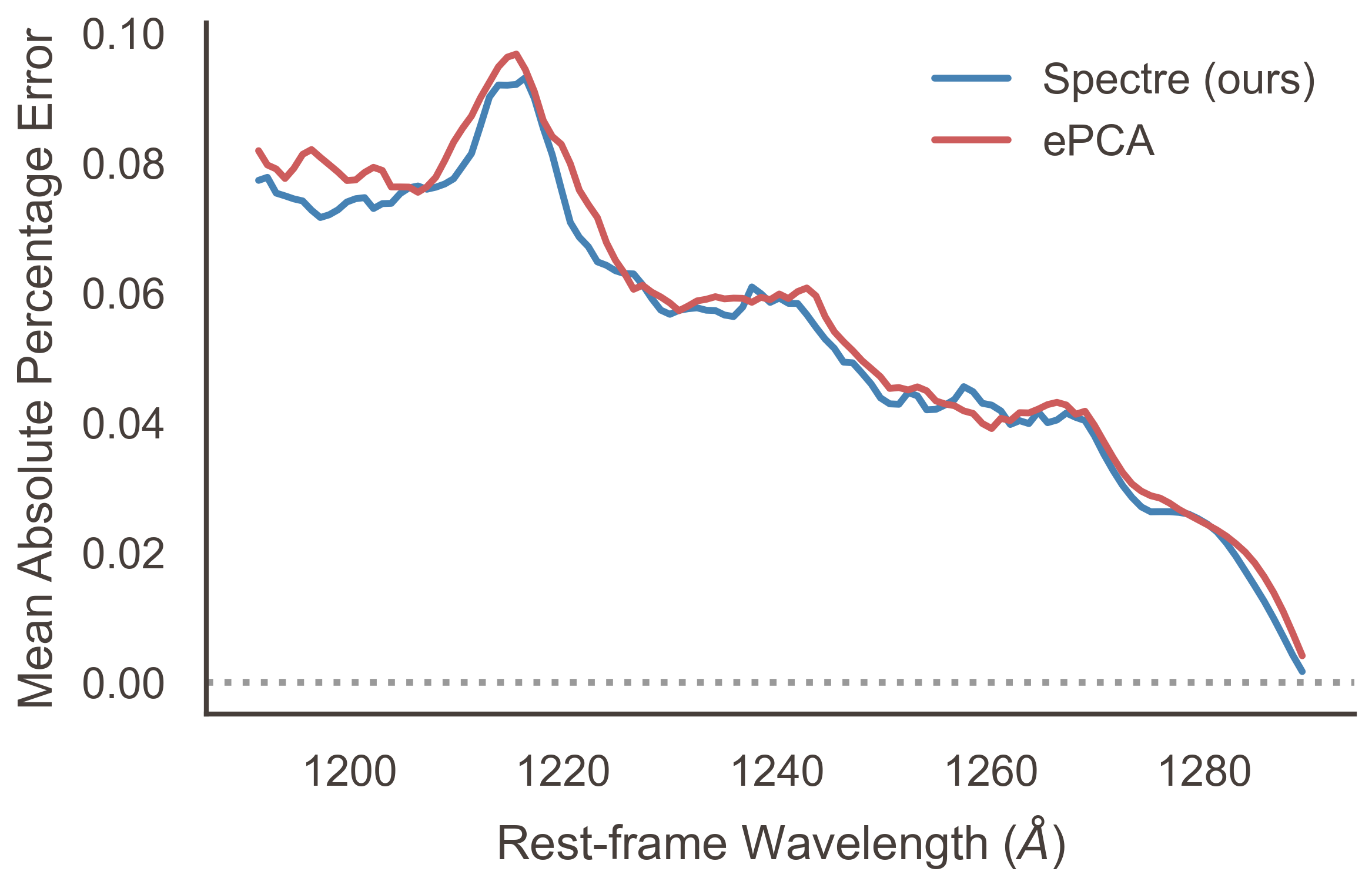}
    \caption{Mean absolute percentage error over the test set as a function of blue-side wavelength for both \spectre\ and the extended PCA (ePCA) method of \citealt{qsanndra} (whose original formulation was introduced in \citealt{daviesPCA}). Error is calculated between the smoothed continua (assumed truth) and \spectre's mean blue-side continua prediction. \spectre\ performs similarly to ePCA while providing an estimation of the full distribution over blue-side continua given the red-side spectrum, allowing for density estimation and sampling (and thereby Monte Carlo estimates of confidence intervals).}
    \label{fig:error_ratio}
\end{figure}

We compare our flow-based model to what is denoted as extended PCA (ePCA): an extension of the original work in \citealt{daviesPCA} presented in \citealt{qsanndra} where PCA is applied independently (in log space) to the blue and red-side spectra and the resulting coefficients related via a linear model. In the original manuscript which describes the use of PCA to predict intrinsic quasar continua \citep{daviesPCA}, the authors chose six and ten components for the blue and red sides, respectively, finding that inclusion of additional components did not yield better results. In \citealt{qsanndra}, this model is extended to include more PCA components---enough to explain 99\% of the variance. The authors find that 36 and 63 principal components are required to meet this criterion on the blue and red side, respectively. Fig.~\ref{fig:error_ratio} shows the mean absolute percentage error of \spectre\ and ePCA as a function of blue-side wavelength. \spectre\ reduces the mean absolute percentage error by ${\sim}5\%$ while offering direct calculation of confidence intervals without ensembling or otherwise. Like other deep learning models, we expect \spectre's performance to improve with dataset size. Near-future surveys such as the Legacy Survey of Space and Time (LSST) at the Vera C. Rubin observatory will provide millions of additional quasar spectra \citep{lsstQuasars} which will likely significantly improve \spectre's performance. 

Additionally, Appendix~\ref{app:add_samples} includes a selection of blue-side continua predictions on the test set labeled with their SDSS designation for reference. These predictions were generated at random and not chosen by hand. More random predictions on the test set can be found at \spectre's GitHub repository: \href{http://www.github.com/davidreiman/spectre}{\ttfamily github.com/davidreiman/spectre}.

\begin{figure*}
	\includegraphics[width=\textwidth]{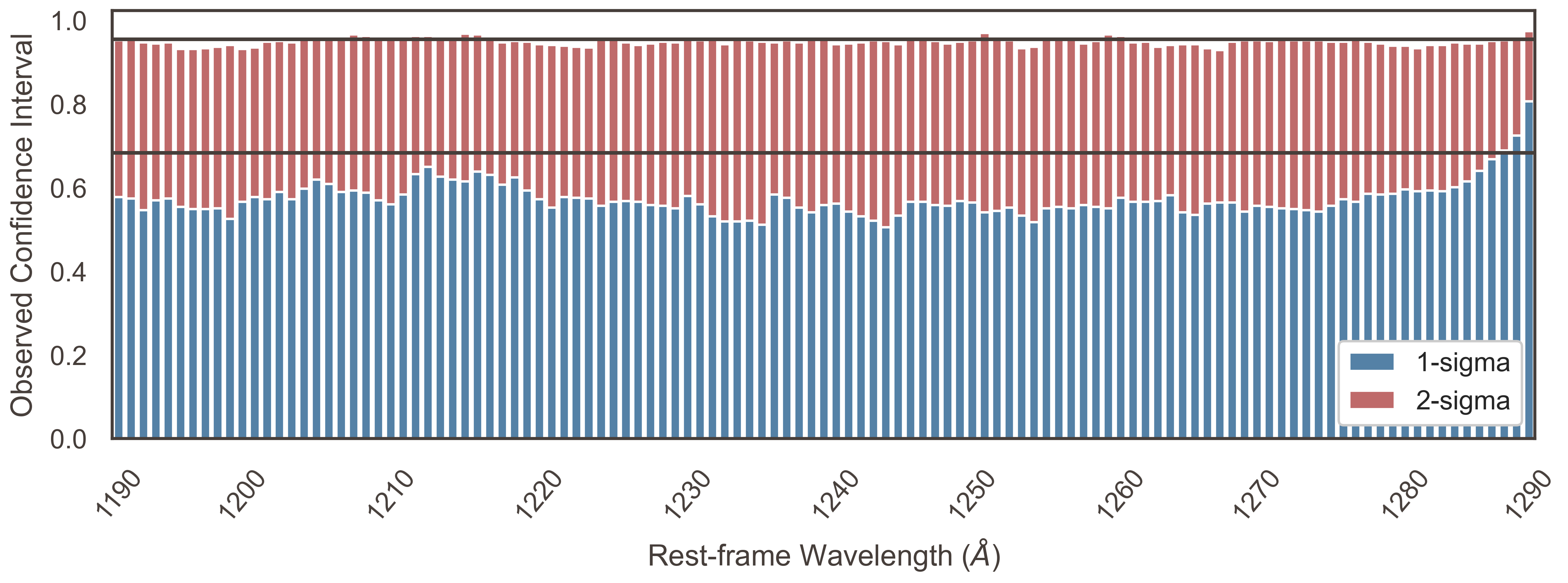}
    \caption{Observed confidence intervals as a function of blue-side wavelength. A calibrated model would make predictions of the marginal density over flux in a given wavelength bin such that $P\%$ of the observed absolute errors fall within the $P\%$ credible interval. We approximate the marginals as Gaussian (which we find to be true to a high degree of accuracy) and show here the observed confidence intervals for 1- and 2-sigma (e.g. $P=68$ and $P=95$). The solid horizontal lines correspond to the expected confidence intervals. We note that \spectre\ tends to be over-confident at the 1-sigma level but is well-calibrated at the 2-sigma level.}
    \label{fig:obs_conf_intervals}
\end{figure*}

\subsection{Uncertainty Assessment}

In this section, we explore the quality of \spectre's uncertainty estimates. We define the pixelwise uncertainty in $N$ random generations from \spectre\ as follows:

\begin{align}
    \sigma_j = \sqrt{\sum_{i=1}^N \frac{(x_{ij} - \bar{x}_j)^2}{N}}
\end{align}
where $i$ indexes the samples generated from \spectre\ and $j$ denotes the pixel or wavelength bin. This makes the assumption that the marginal distribution over flux in each wavelength bin is Gaussian-distributed though in our experiments we find that this is true to a high degree of accuracy.

To measure the calibration of \spectre's uncertainty estimates, we make predictions on all spectra in the test set and compare the observed confidence intervals to the expected confidence intervals. That is, for a calibrated model we would expect to find $P\%$ of the absolute errors within the $P\%$ confidence interval predicted by the model. We can quantify our calibration by checking if this is indeed true. We do so for each pixel (wavelength bin) on the blue-side of all test-set spectra and present the results in Fig.~\ref{fig:obs_conf_intervals}. At the 1-sigma level, we find that on average slightly less than $68\%$ (approximately $57\%$) of the absolute errors lie within \spectre's confidence interval which suggests that our model is slightly over-confident. However, at the 2-sigma level \spectre\ is highly calibrated, as on average \char`\~$95\%$ of the absolute errors fall within \spectre's 2-sigma confidence interval.

\begin{figure}
	\includegraphics[width=\columnwidth]{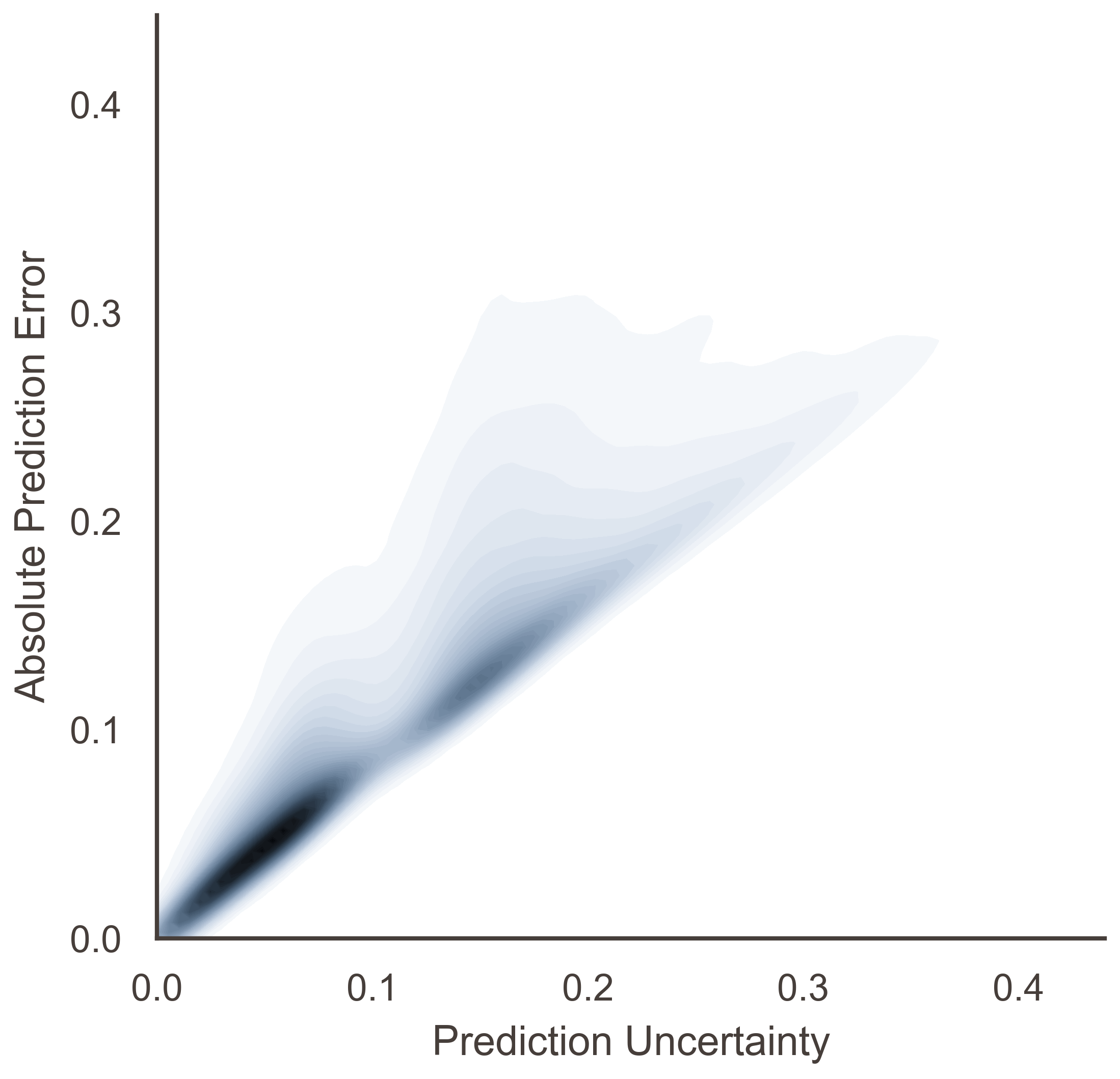}
    \caption{Kernel density estimate of the joint distribution over absolute error and predictive uncertainty in \spectre\ samples over all wavelengths and spectra in the test set.}
    \label{fig:uncertainty_kde}
\end{figure}

We also test the quality of \spectre's uncertainty predictions by computing the joint probability distribution of \spectre's absolute prediction error and predicted uncertainty over all elements of the validation set. The results are presented in Fig.~\ref{fig:uncertainty_kde}. Though we note a sparsely populated tail of underestimated error (an effect also noted in Fig.~\ref{fig:obs_conf_intervals}), \spectre's uncertainty is generally strongly correlated with its own error in its predictions.

There is a growing body of literature on the tuning of an additional hyperparameter, temperature, which modifies the base distribution after training. This is referred to as temperature-scaling and is used to increase the fidelity of model samples and calibrate uncertainties \citep{nn-calibration, img-transformer}. We do not explore these here as this is an active field of research and it is not yet clear which prescription is most reliable \citep{trust-uncertainty}.

\begin{figure*}
  \begin{subfigure}{\textwidth}
    	\includegraphics[width=\textwidth]{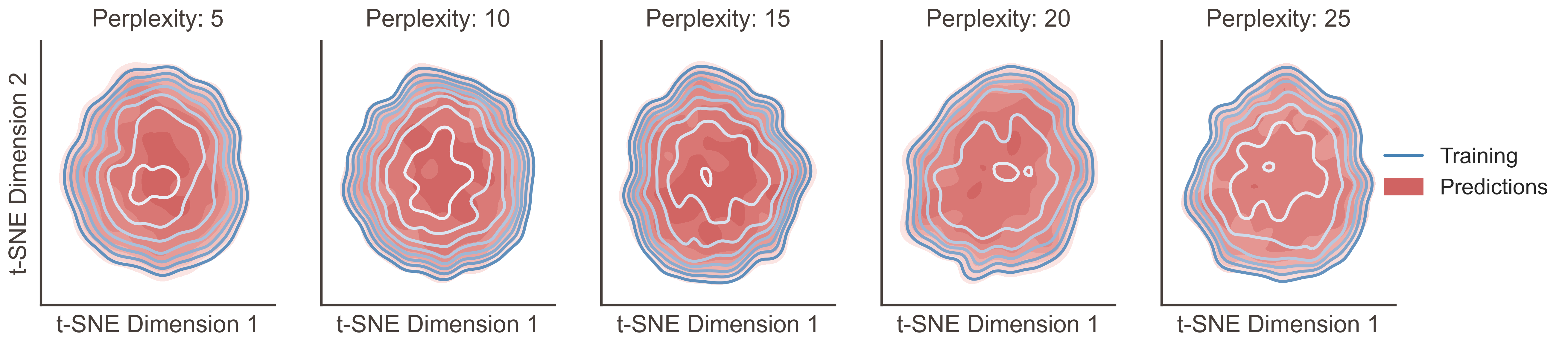}
  \end{subfigure}%
  \vspace*{\floatsep}
  \begin{subfigure}{\textwidth}
	\includegraphics[width=\textwidth]{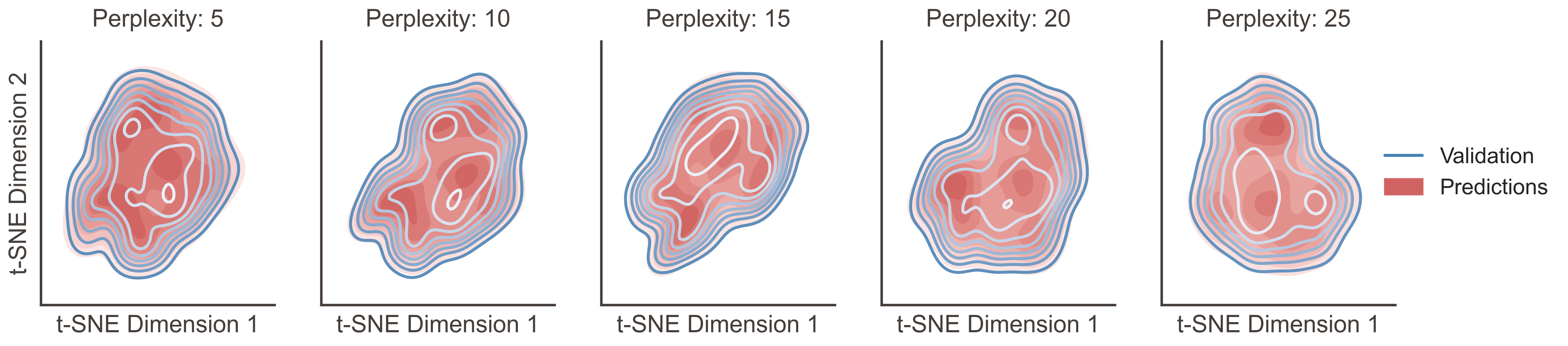}
  \end{subfigure}%
  \vspace*{\floatsep}
  \begin{subfigure}{\textwidth}
	\includegraphics[width=\textwidth]{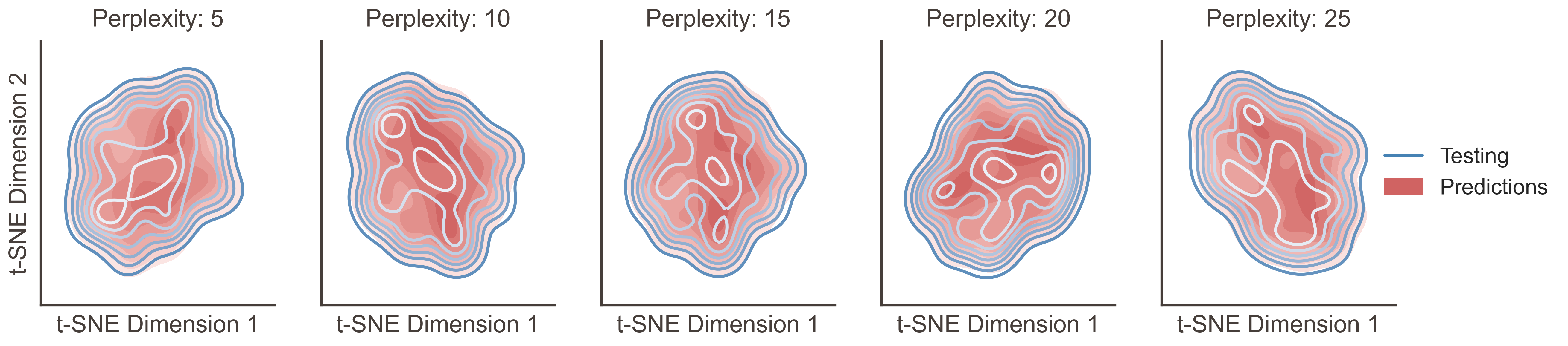}
  \end{subfigure}%
  \caption{Two-dimensional embedding of blue-side spectra produced via t-Distributed Stochastic Neighbor Embedding (t-SNE). \textbf{Top}: Comparisons between our training set continua and associated model predictions. \textbf{Middle}: Comparisons between validation set continua and associated model predictions. \textbf{Bottom}: Comparisons between test set continua and associated model predictions. We display the results for a series of perplexity choices and find that training/validation set distributions consistently overlap with our model samples. Qualitatively, this implies our model's samples accurately cover the full data distribution. }
\label{fig:tsne_combined}
\end{figure*}

\subsection{Sample Coverage}

To ensure that our model achieves full coverage of the training data distribution, we cast the full training and validation sets down to two-dimensional representations with a dimensionality reduction technique known as t-Distributed Stochastic Neighbor Embedding or t-SNE. We then produce a similar number of random samples from \spectre\ and compute the embeddings of these samples for comparison to the training and validation sets. Visualizations of the results are provided in Fig.~\ref{fig:tsne_combined}. Both figures show unique t-SNE embeddings for increasing values of the t-SNE \textit{perplexity}, a hyperparameter which can loosely be interpreted as an initial guess on the number of close neighbors each data point will have. Since the t-SNE algorithm is known to produce very different results for different choices of perplexity, we've shown the results for a variety of choices for completeness. We achieve full coverage of both the training and validation sets. However, we note that the location of the modes are slightly offset though generally overlapping.


\section{Conclusion}
\label{sec:conclusion}

In this manuscript, we have introduced normalizing flows as a powerful and expressive tool for probabilistic modeling in the sciences. Flows boast the ability to perform exact density evaluation, compute uncertainty intervals and carry out one-pass density estimation or sampling (provided one chooses an appropriate flow transform). Many problems in astronomy and beyond can benefit from the use of generative models and such tasks benefit from uncertainty quantification, which other popular models (such as generative adversarial networks) cannot provide.

Among deep generative models, flows are the only models which offer both exact density evaluation and one-pass sampling (provided coupling layer or inverse autoregressive flows are used). Apart from flows, autoregressive models (which factorize high-dimensional joint probability distributions into a product of conditionals via the probability chain rule) are the only other deep generative model capable of exact density evaluation. However, sampling from a D-dimensional distribution with an autoregressive model requires D forward passes since sampling is ancestral.

Flows can also be used to learn priors over data distributions for Bayesian modeling. In maximum a posteriori inference, naive priors are often chosen which don't capture the true complexity of the data at hand. Instead, unconditional flows can provide much more realistic priors on the data given a sizeable dataset. Additionally, flows find use in likelihood-free inference techniques where they are used to approximate the intractable likelihood of a complicated and/or black-box simulator \citep{snl}. This likelihood can then be integrated to obtain the posterior.

Commonly, efficient sampling is the primary model criterion for scientists. Coupling layer or inverse autoregressive flows satisfy this criterion and have recently been used for more efficient sampling in all-purpose numerical integrators \citep{iflow} and in the estimation of the expectation values of physical observables in lattice quantum chromodynamics \citep{qcdFlows}.

In general, flows are capable of density estimation for a wide variety of continuous or discrete-valued data. They have been successfully applied as generative models for images \citep{glow} and text \citep{discreteFlows} and used to perform anomaly detection in particle physics \citep{anode}.

We have applied a specific flow variant---rational quadratic neural spline flows---to the task of intrinsic quasar continua prediction and provided a fully probabilistic model which is readily applicable to current and future high-redshift quasars. Our model consumes the red-side ($\lambda_\text{rest} > \SI{1290}{\angstrom}$) spectrum to estimate a distribution over the blue-side ($\SI{1190}{\angstrom} < \lambda_\text{rest} < \SI{1290}{\angstrom}$) continua. In contrast to previous approaches in the literature, \spectre\ directly models the full probability distribution over blue-side continua and therefore can be resampled arbitrarily many times to generate new plausible blue-side continua and estimate quantities such as confidence intervals without the use of ensembles.

We have also provided two new measurements of the neutral fraction of hydrogen at redshifts $z > 7$. Our results are compatible with reionization constraints from Planck and in agreement with most previous approaches. Our results support a rapid end to ionization however it is difficult to make bold claims on the topic as the available $z > 7$ data is extremely sparse and more robust modeling of the IGM would be prudent. Our modeling of the damping wing makes multiple simplifying assumptions that are untrue: (i) the quasar's proximity zone is entirely ionized, and (ii) the IGM blueward of the proximity zone is uniformly dense and neutral. These concerns can be addressed with future work using full hydrodynamical IGM modeling \citep[e.g.][]{daviesX} in combination with continua predictions from \spectre.

\section*{Acknowledgements}

We acknowledge and thank Daniel Mortlock and Eduardo Ba{\~n}ados for providing the spectra of ULAS J1120+0641 and ULAS J1342+0928, respectively. We would also like to thank Vanessa Boehm,  Kyle Cranmer, Frederick B. Davies, and Brian Maddock for useful comments and discussion.

We acknowledge use of the Lux supercomputer at UC Santa Cruz, funded by NSF MRI grant AST 1828315.

Funding for the Sloan Digital Sky Survey IV has been provided by the Alfred P. Sloan Foundation, the U.S. Department of Energy Office of Science, and the Participating Institutions. SDSS-IV acknowledges
support and resources from the Center for High-Performance Computing at
the University of Utah. The SDSS web site is www.sdss.org.

SDSS-IV is managed by the Astrophysical Research Consortium for the 
Participating Institutions of the SDSS Collaboration including the 
Brazilian Participation Group, the Carnegie Institution for Science, 
Carnegie Mellon University, the Chilean Participation Group, the French Participation Group, Harvard-Smithsonian Center for Astrophysics, 
Instituto de Astrof\'isica de Canarias, The Johns Hopkins University, Kavli Institute for the Physics and Mathematics of the Universe (IPMU) / 
University of Tokyo, the Korean Participation Group, Lawrence Berkeley National Laboratory, 
Leibniz Institut f\"ur Astrophysik Potsdam (AIP),  
Max-Planck-Institut f\"ur Astronomie (MPIA Heidelberg), 
Max-Planck-Institut f\"ur Astrophysik (MPA Garching), 
Max-Planck-Institut f\"ur Extraterrestrische Physik (MPE), 
National Astronomical Observatories of China, New Mexico State University, 
New York University, University of Notre Dame, 
Observat\'ario Nacional / MCTI, The Ohio State University, 
Pennsylvania State University, Shanghai Astronomical Observatory, 
United Kingdom Participation Group,
Universidad Nacional Aut\'onoma de M\'exico, University of Arizona, 
University of Colorado Boulder, University of Oxford, University of Portsmouth, 
University of Utah, University of Virginia, University of Washington, University of Wisconsin, 
Vanderbilt University, and Yale University.




\bibliographystyle{mnras}
\bibliography{mnras_template} 



\pagebreak
\appendix


\section{Likelihood ratio tests}
\label{app:LR}

Here we describe the noise models used when we performed out-of-distribution detection using the likelihood ratio method described in Sec.~\ref{section:LR}. All results are summarized in  Table~\ref{tab:LRsummary}. For reference, the high-z spectra both lie in the 99.9 percentile of in-distribution likelihoods. 

The dataset created with the SDSS noise model is simply composed of the raw flux measurements of quasars whose continua lie in our in-distribution dataset. $\mathcal{N}(0,\ \sigma)$ refers to adding uncorrelated noise sampled from a Gaussian with mean 0 and standard deviation, $\sigma$, to our processed spectra. The noisy PCA model decomposes all continua into PCA components. Uncorrelated noise sampled from Gaussians with variance equal to each components' explained variance is added to each component before reconstructing the continuum.

\begin{table}
	\centering
	\caption{A summary of out-of-distribution detection results on high-z spectra. The lowest likelihood ratio percentiles are shown in bold. We find an out-of-distribution model trained on SDSS flux measurements to provide the best likelihood ratio constraints. }
	\label{tab:LRsummary}
	\begin{adjustbox}{width=\columnwidth}
    \footnotesize
    \renewcommand{\cellalign}{c}
    \begin{tabular}{lcc}
        \toprule
        & \multicolumn{2}{c}{ Likelihood ratio percentile} \\
        Noise model & ULAS J1120+0641 & ULAS J1342+0928 \\
        \midrule
        SDSS & \textbf{61.1} & \textbf{10.4} \\
        $\mathcal{N}(0,\ \sigma=0.07)$ & 61.5  & 17.5  \\
        $\mathcal{N}(0,\ \sigma=0.2)$ & 61.4 & 20.1 \\
        Noisy PCA & 61.4 & 18.8 \\
        \bottomrule
    \end{tabular}
    \end{adjustbox}
\end{table}


\section{Additional Samples}
\label{app:add_samples}

\begin{figure*}
	\includegraphics[width=0.95\textwidth]{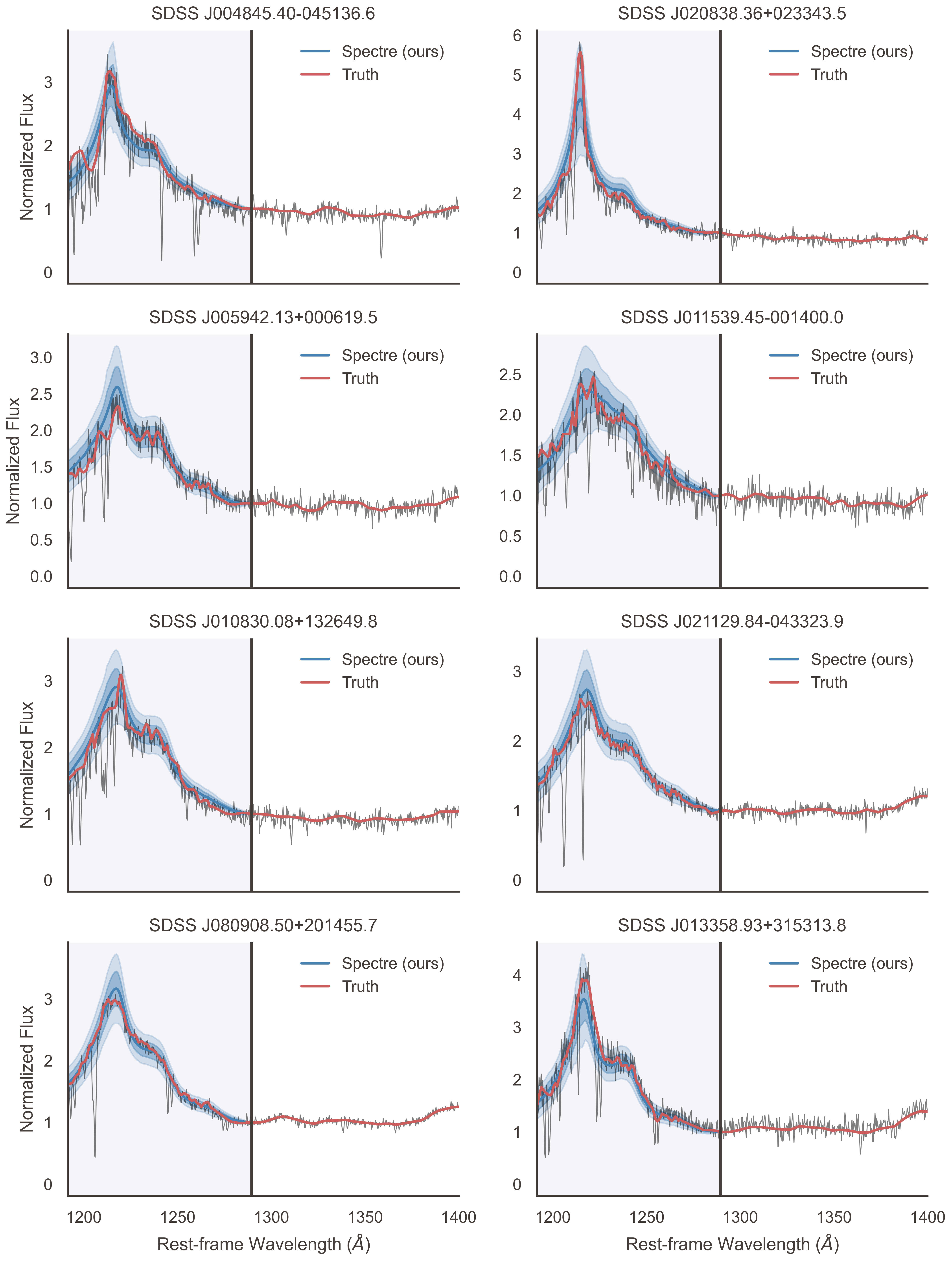}
    \caption{A random selection of eight predictions on moderate redshift test set spectra from eBOSS. \spectre's mean predictions are displayed in blue alongside its 1- and 2-sigma estimates. The assumed truth, shown in red, is the smoothed continua estimation from our preprocessing scheme, and the raw flux is shown in grey. The object in each panel is denoted by its official SDSS designation.}
    \label{fig:random_predictions}
\end{figure*}

We show here (Fig.~\ref{fig:random_predictions}) a random selection of predictions on the test set spectra complete with \spectre's 1- and 2-sigma uncertainties for inspection by the reader. More random predictions are available at \spectre's GitHub repository: \href{http://www.github.com/davidreiman/spectre}{\ttfamily github.com/davidreiman/spectre}.

\section{Model Hyperparameters}
\label{app:model_hyperparams}

In Table~\ref{tab:hyperparameters} we list the hyperparameters of the model used for all experiments and analyses in this manuscript. These hyperparameters were chosen via grid search on a randomly selected validation set of moderate-z quasar spectra from eBOSS.

\begin{table*}
	\centering
	\caption{Model and training hyperparameters, where the leftmost column lists the variable name we use in our codebase (see \href{http://www.github.com/davidreiman/spectre}{\ttfamily github.com/davidreiman/spectre}), the middle column offers a brief description of the hyperparameter's function, and the rightmost column lists the value we used in our final model.}
	\label{tab:hyperparameters}
	
    \begin{adjustbox}{width=\textwidth}
        \footnotesize
        \renewcommand{\cellalign}{c}
        \begin{tabular}{lll}
            \toprule
            Hyperparameter & Description & Value \\
            \midrule
            \verb|n_layers| & Number of coupling layers in flow & 10 \\
            \verb|hidden_units| & Number of hidden units in conditioner & 256 \\
            \verb|n_blocks| & Number of residual blocks in conditioner & 1 \\
            \verb|tail_bound| & (x, y) bounds of spline region & 10.0 \\
            \verb|tails| & Spline function type beyond bounds & linear \\
            \verb|n_bins| & Number of bins in piecewise spline & 5 \\
            \verb|min_bin_height| & Minimum spline bin extent in y & 0.001 \\
            \verb|min_bin_width| & Minimum spline bin extent in x & 0.001 \\
            \verb|min_derivative| & Minimum spline derivative at knots & 0.001 \\
            \verb|dropout| & Dropout probability in flow coupling layers & 0.3 \\
            \verb|use_batch_norm| & Use batch normalization in coupling layers & \verb|True| \\
            \verb|unconditional_transform| & Unconditionally transform identity features & \verb|False| \\
            \verb|use_cnn_encoder| & Use a CNN encoder for redward spectrum & \verb|False| \\
            \verb|encoder_units| & Number of hidden units in encoder & 128 \\
            \verb|n_encoder_layers| & Number of encoder layers & 4 \\
            \verb|encoder_dropout| & Dropout probability in encoder layers & 0.0 \\
            \verb|subsample| & Degree at which to subsample in wavelength & 3 \\
            \verb|log_transform| & Log transform data & \verb|False| \\
            \verb|standardize| & Standardize data by wavelength & \verb|True| \\
            \verb|learning_rate| & Initial learning rate & 5e-04 \\
            \verb|min_learning_rate| & Minimum learning rate & 1e-07 \\
            \verb|anneal_period| & Learning rate annealing period (in batches) & 5000 \\
            \verb|anneal_mult| & Annealing period multiplier after each restart & 2 \\
            \verb|n_restarts| & Total number of warm restarts & 2 \\
            \verb|batch_size| & Batch size during training & 32 \\
            \verb|eval_batch_size| & Batch size during evaluation & 256 \\
            \verb|eval_n_samples| & Number of samples to draw from flow during evaluation & 1000 \\
            \verb|grad_clip| & Maximum gradient norm during training & 5.0 \\
            \verb|n_epochs| & Maximum number of training epochs & 200\\
            \bottomrule
        \end{tabular}
    \end{adjustbox}
\end{table*}

\bsp	
\label{lastpage}
\end{document}